\documentclass[useAMS,usenatbib]{mnras}

\usepackage[T1]{fontenc}
\usepackage{ae,aecompl}
\usepackage{graphicx}
\usepackage{amsmath,amssymb}
\usepackage{color,ulem}

\definecolor{webgreen}{rgb}{0,.5,0}
\definecolor{webbrown}{rgb}{.6,0,0}

\hypersetup{%
   colorlinks=true,%
   breaklinks=true,%
   plainpages=false, bookmarksnumbered, bookmarksopen=true,
   bookmarksopenlevel=1,%
   urlcolor=webbrown, linkcolor=blue, citecolor=webgreen,
   }
   
\newcommand       \ba           {\begin{eqnarray}}
\newcommand       \ea           {\end{eqnarray}}
\def\bm#1{\mbox{\boldmath $#1$}}
\def \be{\begin{equation}}
\def \ee{\end{equation}}

\title[Global stability analysis and non-linear simulations of thermal instability]{Cold gas in cluster cores: Global stability analysis and non-linear simulations of thermal instability}  

\author[P. P. Choudhury, P.\ Sharma]
{Prakriti Pal Choudhury $^\dag$, Prateek Sharma$^\ddag$\\
$^\dag$ Department of Physics, Indian Institute of Science, Bangalore , India 560012 (prakriti@physics.iisc.ernet.in)\\
$^\ddag$Department of Physics and Joint Astronomy Program, Indian Institute of Science, Bangalore, India 560012 (prateek@physics.iisc.ernet.in)}

\begin{document}
\maketitle

\label{firstpage}

\begin{abstract}
We perform global linear stability analysis and idealized numerical simulations in global thermal balance to understand the condensation of  cold gas from hot/virial atmospheres (coronae), in particular the intracluster medium (ICM). 
We pay particular attention to geometry (e.g., spherical versus plane-parallel) 
and the nature of the gravitational potential. Global linear analysis gives a similar value for the fastest growing thermal instability modes in spherical 
and Cartesian geometries. Simulations and observations suggest that cooling 
in halos critically depends on the ratio of the cooling time to the free-fall time ($t_{\rm cool}/t_{\rm ff}$). Extended cold gas condenses out of the ICM 
only if this ratio is smaller than a threshold value close to 10.
Previous works highlighted the difference between the nature of cold gas condensation in spherical and plane-parallel
atmospheres; namely, cold gas condensation  
appeared easier in spherical atmospheres. This apparent difference due to geometry arises because the 
previous  plane-parallel simulations focussed on 
{\it in situ} condensation of multiphase gas but spherical simulations studied condensation {\it anywhere} in the box. Unlike previous claims, 
our nonlinear simulations show that there are only minor differences in cold gas condensation, either in situ or anywhere, for different geometries. 
The amount of cold gas depends on the shape of $t_{\rm cool}/t_{\rm ff}$; gas has more time to condense if gravitational acceleration decreases toward the center.
In our idealized plane-parallel simulations with heating balancing cooling in each layer, there can be significant mass/energy/momentum transfer across layers 
that can trigger condensation and drive $t_{\rm cool}/t_{\rm ff}$ far beyond the critical value close to 10. 
\end{abstract}


\section{Introduction}           
\label{sect:intro}
Galaxy formation in dark matter haloes is understood to be a consequence of the combined effects of cooling, heating, and gravity 
(e.g., see \citealt{1991ApJ...379...52W,2005Natur.435..629S,2006MNRAS.370..645B}). Gas does not simply cool
out of the hot virialized intracluster medium (ICM), but is maintained in rough thermal balance due to energy input by active galactic nucleus (AGN) 
jets in cores of galaxy clusters (\citealt{mcnamara2007}). 

Observations show that the galaxy cluster cores with short cooling times or small core entropies have spatially extended multiphase gas and current star formation 
(\citealt{Raf2008,accept09,mcdonald10}). Recent theoretical
and computational works, beginning with idealized simulations of local thermal instability in global thermal balance 
(\citealt{mccourt12,sharma12}) and culminating with the simulations of feedback-driven AGN jets interacting with the ICM (\citealt{gaspari12,li14,prasad15}),
have shown that cold gas condenses out of the ICM if the ratio of cooling time to the free-fall time ($t_{\rm cool}/t_{\rm ff}$) is $\lesssim 10$. Moreover, cool 
cluster cores with jet feedback show heating and cooling cycles in which this ratio (min[$t_{\rm cool}/t_{\rm ff}$]) varies from a few to few tens 
(\citealt{prasad15,li2015}). A rough floor of $t_{\rm cool}/t_{\rm ff}\approx 10$ has been recently confirmed in observations of halos with temperatures 
ranging from 0.5 
to 15 keV (\citealt{Voitnat2015}). The interpretation is that AGN feedback, fueled by condensing cold gas, kicks in for a shorter cooling 
time and reheats the hot phase such that $t_{\rm cool}/t_{\rm ff} \gtrsim 10$ everywhere.

Idealized simulations of hot ICM in global thermal balance, confined by 
the dark matter potential, have shown that the initial ratio of cooling time to free-fall time has to be less than a critical value (close to 10) if {\it extended}
cold gas has to condense out of the hot ICM (\citealt{sharma12}); cold gas does not appear for a larger $t_{\rm cool}/t_{\rm ff}$ 
even if we wait for a long time.
In contrast, a recent study based on idealized simulations, with a constant ratio of cooling-time to free-fall time throughout, 
suggests that the initial $t_{\rm cool}/t_{\rm ff}$ determines the onset time of cold gas formation but cold gas condenses eventually if one waits long 
enough (\citealt{meece15}).  We 
critically examine this issue in our paper, and show that the authors' claim is a trivial consequence of gravity vanishing in the midplane 
of their simulations. In other words, gravitational acceleration vanishes at 
the midplane, and hence gravity cannot suppress cold gas condensation there. 
As we show later, fluctuations generated due to local thermal instability at the midplane can lead to cold gas far from there. 

\citet{meece15} also highlighted that there is not much dependence on geometry (e.g., plane-parallel versus spherical) of cold gas condensation from 
virial atmospheres. Prior to this, it was believed that it is much easier for cold gas to condense in spherical geometry as compared to a plane-parallel atmosphere. 
Multiphase gas was claimed to condense out if min$(t_{\rm cool}/t_{\rm ff}) \lesssim 10$ (more precisely, $t_{\rm TI}/t_{\rm ff}$,\footnote{While we can calculate 
$t_{\rm cool} \equiv (3/2)nk_BT/n_en_i \Lambda$ and $t_{\rm TI} \equiv (10/9) t_{\rm cool}$ (for a density/temperature independent constant heating rate; 
$[10/3] t_{\rm cool}$ for a local heating rate linearly proportional to density; for details, see \citealt{mccourt12}) in simulations, we do not know $t_{\rm TI}$ in 
cluster cores because we do not know the microscopic heating mechanism. Simulations with turbulent heating/mixing give a growth rate consistent with 
a heating rate density independent of density, as required to match observations (\citealt{banerjee14}).} the ratio of local thermal 
instability growth-time and free-fall time) in spherical geometry (\citealt{sharma12}) and $ \lesssim 1$ in a plane-parallel atmosphere (\citealt{mccourt12}).
This anomaly of a factor of $10$ has been argued to be the effect of geometric compression in spherical geometry, which makes it easier to form cold gas. 
Experiments with a phenomenological compression term to simulate the geometric effect in a simple blob model (motivated by \citealt{ps05}), have shown 
that the formation of cold gas may be 
triggered for $t_{\rm TI}/t_{\rm ff}\lesssim 10$ (\citealt{ashmeet14}). Our reanalysis of numerical simulations confirms that geometry %indeed 
does not play an important role in predicting the onset of cold gas condensation. 
 \citet{sharma12} and  \citet{mccourt12} found a dependence on geometry because the former concentrated on condensation
{\it anywhere} in the computational domain, whereas latter focussed on {\it in situ} condensation ( i.e., condensation in the plane where $t_{\rm TI}/t_{\rm ff}$
is minimum) of multiphase gas.

Given the importance of cold gas condensation in hot coronae, in this paper we combine global linear stability analysis and idealized numerical simulations 
to have a definitive say on the properties of the {\it extended} cold gas condensing out of the hot atmosphere. 
The paper is divided into two main parts: a global linear stability analysis and numerical simulations of local thermal instability in global thermal balance. 
Section \ref{sec:phy_set} presents our physical setup and the equations that we solve. 
Section \ref{sec:global_linear} describes the global linear stability analysis and section \ref{sec:num_sims} is on results from numerical simulations. We conclude 
in section \ref{sec:conc}.

\section{Physical setup}

In this section we describe the physical setup that we use to study the evolution of local thermal instability in an atmosphere satisfying global 
thermal equilibrium. This is a useful model to understand the origin of multiphase gas in galaxy cluster cores. The same basic setup is 
used for the global thermal instability analysis in section \ref{sec:global_linear} and for idealized numerical simulations in section 
\ref{sec:num_sims}.

\label{sec:phy_set}
\subsection{Model \& equations}
\label{sec:sec1}
The background/unperturbed ICM is in hydrostatic and thermal equilibrium. 
The equations governing the evolution of gas in the ICM are 
\ba
\label{eq:eq1}
\frac{D \rho}{Dt} &=& - \rho {\bf \nabla \cdot \bm{v}}, \\
\label{eq:eq2}
\frac{D \bm{v}}{Dt} &=& -\frac{1}{\rho} {\bf \nabla} p - g \bm{\hat{r}}, \\
\label{eq:eq3}
\frac{p}{(\gamma-1)} \frac{D}{Dt} \left [ \ln \left ( \frac{p}{\rho^\gamma} \right ) \right ] &=& -q^-(n,T) + q^+(r,t), 
\ea
where $D/Dt$ is the Lagrangian derivative, $\rho$, $\bm{v}$ and $p$ are mass density, velocity and pressure; $\gamma=5/3$ is the adiabatic index; 
$q^-(n,T) \equiv n_e n_i \Lambda(T)$ 
($n_e \equiv \rho/[\mu_e m_p]$ and $n_i \equiv \rho/[\mu_i m_p] $ are electron and ion number densities, respectively; $\mu_e=1.17$, $\mu_i=1.32$, 
and $m_p$ is proton mass) 
and $q^+(r,t) \equiv \langle q^- \rangle$ (this ansatz imposes thermal balance in shells), $\Lambda(T)$ is the temperature-dependent cooling function, 
and angular brackets indicate shell averaging. The background hydrostatic 
equilibrium implies $dp_0/dr = -\rho_0 g$, where a subscript `0' refers to equilibrium quantities and acceleration due to gravity 
$g \equiv d\Phi/dr$ ($\Phi$ is the fixed gravitational potential). 
The background quantities are functions of $r$ and 
the perturbations, in general, depend on all coordinates and time.

Our global linear stability analysis in section \ref{sec:global_linear} is solved as a linear eigenvalue problem in radius. All perturbed quantities (density, 
velocity, etc.) are expanded in a  Chebyshev polynomial basis, and the matrix equation for eigenvalues and eigenfunctions is solved numerically. For 
numerical simulations in section \ref{sec:num_sims} we simply initialize isobaric density perturbations and study their evolution; strict thermal equilibrium
in shells ($q^+=\langle q^-\rangle$) is maintained at all times.

\subsection{Gravitational acceleration profiles}

For most of our runs we use the NFW profile for the potential and gravitational acceleration, 
\ba
\label{eq:phi_NFW}
 \Phi_{\rm NFW}  &=& -\frac{GM_{200} \ln (1+c_{200}r/r_{200})}{ [\ln (1+c_{200}) - c_{200}/(1+c_{200}) ] r}, \\
\label{eq:g_NFW}
g_{\rm NFW} &=& \frac{d}{dr} \Phi_{\rm NFW},
\ea
where $M_{200} = 5.24 \times 10^{14} M_\odot$ is the dark matter halo mass within $r_{200}$ and $r_{200}$ is the approximate 
virial radius (\citealt{nfw96}), within which the mean density is 200 times the critical density of the universe ($9.2 \times 10^{-30}$ g cm$^{-3}$), and 
$c_{200}=3.3$ is the concentration parameter. While the NFW potential is normally used only in spherical geometry, we also use it in Cartesian/plane-parallel 
setups in which the gravitational potential and acceleration have the same dependence on $|z|$. This is done to compare cold gas condensation in spherical
and plane-parallel geometries.

In order to compare with previous idealized simulations (e.g., \citealt{mccourt12,meece15}), especially
the influence of the form of gravity on cold gas condensation, we also perform runs with the following gravitational accelerations 
\ba
\label{eq:g_MOV}
&& g_{\rm MOV}(r) = g_0 \tanh \left ( \frac{r}{r_s} \right ),   \\
\label{eq:g_MSQP}
&& g_{\rm MSQP}(r) = g_0 \frac{r/r_s}{(1 + [r/r_s]^2)^{\frac{1}{2}}}, 
\ea
where $r_s$ is a scale height (chosen 5 kpc for MSQP and 50 kpc for MOV; for the same $r_s$, the shapes of the two accelerations are very similar). 
Note that $r$ is simply replaced by $z$ for plane-parallel atmospheres.

\begin{figure}
 \includegraphics[width=.5\textwidth]{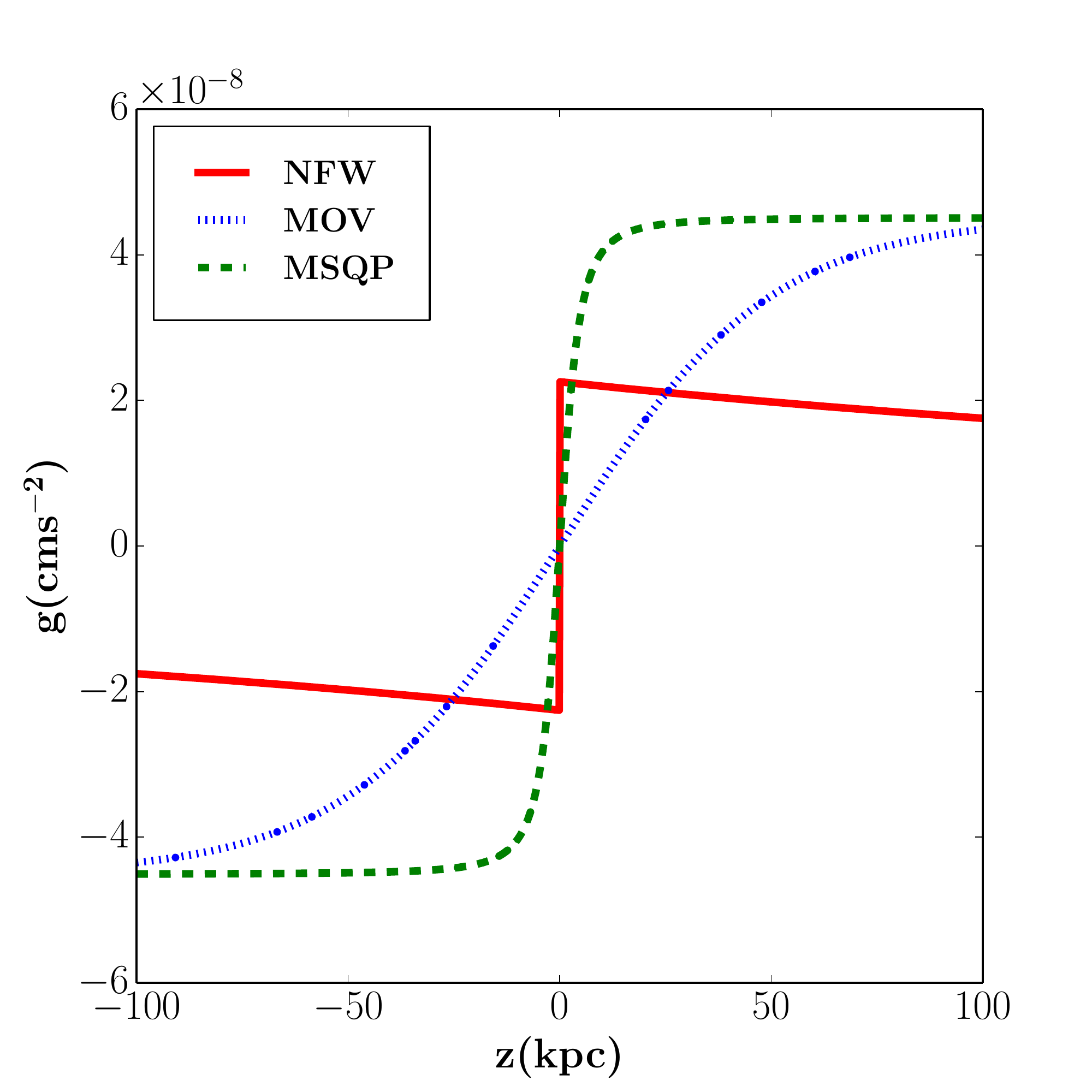}
 \caption{Different gravitational acceleration profiles that we use (Eqs. \ref{eq:g_NFW}-\ref{eq:g_MSQP}) as a function of distance from the midplane (or the center, 
 in case of spherical geometry).  }\label{fig:fig01}
\end{figure}

In order to make meaningful comparisons with the realistic NFW profiles,
we take the value of $g_0$ such that the order of magnitude of gravitational acceleration is similar to that of the realistic NFW gravity. 
We also carry out idealzed MSQP numerical simulations in which $g_0=1$.
Although the magnitude of gravitational accelerations is similar, their spatial dependence is different; in particular, gravity becomes 
weaker toward the center for MSQP and MOV gravities (Fig. \ref{fig:fig01}). Note that at small radii $g \propto r$ for MOV and MSQP gravities, and thus
$t_{\rm ff} = (2r/g)^{1/2}$ is approximately a constant and $t_{\rm cool}/t_{\rm ff}$ peaks at the center because of the highest density at the center. In 
contrast, $t_{\rm cool}/t_{\rm ff}$ peaks at $\sim 10$ kpc for a realistic NFW gravity (c.f. Fig. \ref{fig:fig07}).

\subsection{Equilibrium profile}
\label{sec:sec2}
The entropy profile of the ICM in initial hydrostatic equilibrium is modeled by
\be
\label{eq:ent}
K(r) = \frac{T_{\rm keV}}{n_e^{\gamma - 1}} = K_0 + K_{100}\left(\frac{r}{r_{100}}\right)^{\alpha},
\ee
where $r_{100} = 100 \ \rm  kpc$; we choose  $K_{100} = 80$ keV cm$^2$ for all runs and linear analysis (\citealt{accept09}). The core entropy value
$K_0$ is varied to obtain various $t_{\rm TI}/t_{\rm ff}$ profiles in numerical simulations. In global linear stability analysis (section \ref{sec:global_linear}), 
we choose $K_0=8$ keV cm$^2$ for NFW gravity and $K_0=23$ keV cm$^2$ for MOV gravity (this choice gives a similar min[$t_{\rm TI}/t_{\rm ff}]\approx 7.3$ ). 
The equilibrium profiles are obtained as a function of the radial distance from the center ($r$; or distance from the midplane $[z]$ in case of a plane-parallel 
atmosphere) by assuming hydrostatic equilibrium in the external potential well (NFW, MSQP, or MOV). The electron number density at 100 kpc
is $0.00875$ cm$^{-3}$. The outer ICM electron density combined with the entropy profile and hydrostatic equilibrium completely specify the ICM profiles.

In addition to the MOV and MSQP gravitational field with realistic entropy profiles (Eq. \ref{eq:ent}), we carry out some idealized MSQP 
(isothermal and isentropic, with profiles given by Eqs. 9 \& 10 of \citealt{mccourt12}; see Table \ref{table:sims}) runs to compare with previous simulations.

\subsection{Idealized heating and cooling}
We use a fit to the plasma cooling function with a third of the solar metallicity, given by Eq. 12 and the solid line in Fig. 1 of \citet{sharma2010}; idealized MSQP
runs use a cooling function $\Lambda = \Lambda_0 T^{1/2}$ (see Table \ref{table:sims}). Since
the cluster core plasma appears to be in rough thermal balance, we adopt an {\it idealized} heating function such that $r/z$-dependent heating rate 
at every radius/height and at all times is equal to the average cooling rate at the 
same height; i.e., $q^+ \equiv \langle n_e n_i \Lambda (T)\rangle $, where angular brackets 
represent shell averaging. Note that our heating rate per unit volume is independent of the local density and temperature.

In the plane-parallel numerical simulations, we have excluded heating and cooling in the region $-1~{\rm kpc} \leq z \leq 1~{\rm  kpc}$ because gravity necessarily 
vanishes at the midplane, where cooling overdense blobs are not %mixed due to 
affected by gravity, and local thermal instability inevitably leads to multiphase gas within a few cooling 
times, irrespective of $t_{\rm TI}/t_{\rm ff}$.\footnote{We do not have to  turn off cooling and heating within 1 kpc in spherical ICM runs because the 
central region is excised for these.} Moreover, we are mainly  
concerned about {\it extended} cold gas filaments observed in cool cluster cores. As we note in section \ref{sec:midplane_cooling}, allowing thermal-instability 
generated multiphase gas in the midplane can artificially enhance 
condensation beyond the midplane even in atmospheres with large $t_{\rm TI}/t_{\rm ff}$.

\subsection{Important timescales}
In this section we define the important timescales relevant to our setup of local thermal instability in global thermal equilibrium. 
Since $t_{\rm cool}/t_{ \rm ff}$ or $t_{\rm TI}/t_{ \rm ff}$ plays a crucial role
in cold gas formation, we define the thermal instability (TI) time-scale (the inverse of the local exponential growth rate for a constant 
heating rate per unit volume)
\be
\label{eq:tTI}
t_{\rm TI} = \frac{\gamma t_{\rm cool}}{(2 - d \ln \Lambda/d \ln T)},
\ee
where
\be
\label{eq:tcool}
t_{\rm cool} = \frac{nk_BT}{(\gamma - 1)n_e n_i \Lambda}.
\ee
For free-free cooling (with $\Lambda \propto T^{\frac{1}{2}}$) relevant to clusters, $t_{\rm TI} = (10/9)t_{\rm cool}$. The free-fall time
\be 
\label{eq:tuff}
t_{\rm ff} = \left( \frac{2r}{g} \right )^{\frac{1}{2}},
\ee
where $g(r)$ ($r$ is replaced by $z$ for plane-parallel atmospheres) is the gravitational acceleration at the radius of interest.

\section{Linear stability analysis: Global modes}
\label{sec:global_linear}

\begin{table*}
 \caption{Global linear stability analysis models with axisymmetric perturbations}
{\centering
\begin{tabular}{c c c c c c c}
\hline\hline
Gravity & Parameters & $K_0$ & ${\rm min}(t_{\rm TI}/t_{\rm ff})$ & ${\rm min}(\sigma_{\rm real}^{-1})$& $t_{\rm TI}$ at ${\rm min}(t_{\rm TI}/t_{\rm ff})$ & $t_{\rm TI}$ at $N_{\rm max}$ \\
 & & $({\rm keV~cm^2})$ & & (Gyr) & (Gyr) & (Gyr)  \\
\hline
 spherical, Cartesian, NFW & $k_x r_{\rm out} = 100$ & $8$ & $ 7.3$ & 0.57 & $ 0.36$  & $ 0.22$ \\
 global spherical harmonics, NFW & $l=10$ & $8$ & $7.3$ & 0.446   & $ 0.36$  & $ 0.22$  \\
 global spherical harmonics, NFW & $l=2$ & $8$ & $7.3$ & 0.448   & $ 0.36$  & $ 0.22$  \\
 spherical, Cartesian, MOV &$k_x r_{\rm out} = 100$ & $23$ & $7.3$ & 3.8  & $ 0.61$  & $ 1.84$  \\
\hline
\end{tabular}
}
\textbf{Notes:} In all cases, the fastest growth timescale ($\sigma_{\rm real}^{-1}$) for resolved modes is longer than the 
local thermal instability timescale.
 \label{table:linanalysis}
\end{table*}

While the local WKB analysis of thermal instability in global thermal and dynamical equilibrium is well-known (\citealt{field65}), 
we carry out a global linear analysis 
of our equilibrium set up, focusing on the growing (strictly speaking, overstable, which is both growing and oscillating) thermal instability modes. 
Linear analysis of 
thermal instability in cluster cooling flows has a long history (e.g., \citealt{malagoli87,balbus88,balbus89,kim03}).  For an equilibrium with a
background flow, Lagrangian analysis appears more reliable (\citealt{balbus88,balbus89}) but as we explain in the next paragraph, a static 
background is a better description of cool cluster cores. For the cluster cores in rough thermal balance there is no background flow and the Eulerian 
and Lagrangian approaches (for both local and global analyses) are equivalent (see the discussion after Eq. 31 in \citealt{kim03}).

Recent results have shed
new light on thermal equilibrium in galaxy cluster cores, in particular the role of AGN feedback in providing rough {\it global} thermal balance and the lack of 
massive cooling flows. Moreover, now it is clear that the anisotropic nature of thermal conduction prevents it from playing a dominant role in global thermal 
balance (\citealt{parrish09,wagh14} and references therein). Therefore, in this paper we study the nature of global thermal instability with a background equilibrium 
(both thermal and hydrostatic) state which is consistent with these recent advances; i.e., we neglect thermal conduction and an equilibrium inflow
(these assumptions make the analysis more tractable; the observed multiphase inflow is small and is due to the nonlinear evolution of local thermal instability), 
and assume global thermal balance ($q^+=\langle q^- \rangle$).

One of the reasons we carry out a global linear instability analysis in spherical geometry and compare it with the same analysis in a plane-parallel atmosphere
is to see if the {\it apparent}\footnote{As we show later, using numerical simulations, that there is not much difference between condensation of cold gas in 
spherical and plane-parallel atmospheres. This is also pointed out by \citet{meece15}. This confusion arose as the plane-parallel simulations of 
\citet{mccourt12} focused on {\it in situ} condensation of 
cold gas, whereas the spherical simulations of \citet{sharma12} concerned with condensation {\it anywhere} outside of 1 kpc, not necessarily in situ at 
$\sim10$ kpc where $t_{\rm TI}/t_{\rm ff}$ is minimum.}  result, that it is easier for cold gas to condense out of the hot ICM in spherical geometry compared to a 
plane-parallel geometry, can be due to a higher growth rate (equivalently, a shorter growth time) in spherical geometry because of geometrical compression 
as an over dense blob falls in toward the center. The evolution of overdensity, assuming a linear growth timescale $t_{\rm TI}$ and a nonlinear mixing 
time $\tau_{\rm nl}$,  is roughly described by 
$$
\frac{d \delta}{dt} \approx \frac{ \delta}{t_{\rm TI}} - \frac{\delta^2}{\tau_{\rm nl}},
$$
which for the amplitude in saturated state gives $\delta_{\rm sat} \approx \tau_{\rm nl}/t_{\rm TI}$. Thus, the saturated amplitude is inversely 
proportional to linear growth time.  A shorter $t_{\rm TI}$ can, in principle, lead to higher overdensities and  easier condensation of cold gas in 
spherical geometry. 

In fact, based on this anticipation, \citet{ashmeet14} erroneously (it appears in light of the global results in this paper) introduced a linear term ($-2g/r$ 
term in Eq. 24 of their paper) that leads to a larger 
instability rate in spherical geometry compared to a plane-parallel atmosphere. Our global linear instability analysis can easily reveal any difference in the growth 
rate of thermally unstable local and global modes. Our 
results show that there is no substantial difference between the linear growth rates in Cartesian 
and spherical atmospheres (c. f. Fig. \ref{fig:fig06}). This is consistent with nonlinear simulations of section \ref{sec:num_sims}, which show only minor 
differences between the threshold $t_{\rm TI}/t_{\rm ff}$ for cold gas condensation in spherical and plane-parallel atmospheres.

We assume that the time dependence of all perturbations (denoted by a subscript 1) is given by $e^{\sigma t}$, where $\sigma$ is the growth rate of the mode 
(it is an imaginary number 
for purely oscillating modes). The linearized equations for perturbations (in spherical coordinates; generalization to a plane parallel atmosphere 
is straightforward) are
\ba
\label{eq:eq4}
\nonumber
 \sigma \rho_1 &+&  \frac{1}{r^2} \frac{\partial}{\partial r}  ( r^2 \rho_0 v_{r1} ) + \frac{\rho_0}{r \sin \theta} \frac{\partial}{\partial \theta} ( \sin \theta v_{\theta 1}) \\
&+& \frac{\rho_0}{r \sin \theta} \frac{\partial}{\partial \phi} v_{\phi 1}    =   0, \\
\label{eq:eq5}
 \sigma \rho_0 v_{r1} &=& - \frac{\partial p_1}{\partial r} - \rho_1 g, \\
\label{eq:eq6}
 \sigma \rho_0 v_{\theta 1} &=& -\frac{1}{r} \frac{\partial p_1}{\partial \theta}, \\
\label{eq:eq7}
 \sigma \rho_0 v_{\phi 1} &=& -\frac{1}{r \sin \theta} \frac{\partial p_1}{\partial \phi}, \\
\label{eq:eq8}
 \sigma \frac{s_1}{s_0}  &+&  \frac{\gamma N^2 v_{r1}}{g} = \frac{-1}{t_{\rm cool 0}} \left [ 2 \frac{\rho_1}{\rho_0} + \frac{d \ln \Lambda}{d \ln T} \frac{T_1}{T_0}  \right ]
\ea
where $T_1/T_0 = p_1/p_0 - \rho_1/\rho_0$, 
\be
\label{eq:BV}
N^2 \equiv \frac{g}{\gamma} \frac{d}{dr}\ln \left ( \frac{p_0}{\rho_0^\gamma} \right )
\ee
 is the local Brunt-V\"ais\"al\"a (BV) frequency, and 
$t_{\rm cool0} \equiv p_0/\{[\gamma-1] n_0^2 \Lambda(T_0) \}$ is the local cooling time. We have also defined a pseudo-entropy as
 $s \equiv p /\rho^\gamma$ such that 
$s_1/s_0 = p_1/p_0 - \gamma \rho_1/\rho_0$.  The expression on the right hand side of Eq. \ref{eq:eq8} 
assumes that $q^+_1=0$. The perturbed heating rate density $q^+_1$ is zero if $\partial/\partial \theta$ or $\partial/\partial \phi$ is non-zero (from our 
assumption of thermal balance in shells). Otherwise, $q^+_1$ 
depends on our assumption: $q^+_1=q^-_1$ if we assume thermal balance in shells at all times (even for perturbations; 
as in our simulations discussed 
in section \ref{sec:num_sims}); $q^+_1=0$ is also a valid choice as  thermal balance is only required to be maintained in equilibrium.
%Note that the perturbed heat loss function has the form on the right hand side of Eq. \ref{eq:eq8} only when 
%$\partial/\partial \theta$ or $\partial/\partial \phi$ is non-zero. Otherwise, $q^+_1 = \langle q^-(n,T) \rangle_1 = q^-_1$; i.e., perturbed fluid elements 
%are in exact thermal balance. 
%{\bf{Precisely, if we take a one-dimensional system with only radial direction, there will be no local thermal instability because
%we are balancing heating and cooling at each radius, just like in our simulations, to make an exact comparison. However, it is worth noting that this assures a global thermal 
%equilibrium, that is, the background medium is always in thermal balance, similar to the situation in all our simulations.}}

Eqs. \ref{eq:eq4}-\ref{eq:eq8} are in the form of an eigenvalue problem. In absence of cooling and heating, these equations describe the coupled sound and 
internal gravity waves.\footnote{In absence of gravity, the only direction in the homogeneous isotropic equilibrium is along the wavenumber $\bm{k}$. 
Thus, the modes in this case are the two oppositely propagating sound waves, and one entropy mode with $\sigma=0$ and only non-zero density
 and temperature perturbations. With gravity, $\bm{g}$ and $\bm{k}$ define a plane ($x-z$ plane, assuming $\bm{k} \cdot \bm{g} \ne 0$) 
 and we obtain two internal gravity modes in addition to sound waves.} These waves are essentially decoupled when the two frequencies are 
 very different (i.e., in the usual Boussinesq approximation).
With cooling and heating, both the internal gravity waves and sound waves can have growing or damped solutions, depending on the perturbed generalized 
heat loss function $q^{-}-q^+$ (Eq. \ref{eq:eq3}). For local modes with short sound crossing times compared to the cooling time
the perturbation is usually evaluated in isobaric (isentropic) approximation for internal gravity waves (sound waves; 
\citealt{field65}). For our cooling function ($\Lambda[T]$), only the internal gravity waves show growing modes. Since our method does not make 
Boussinesq approximation, our global method calculates exact oscillation frequencies/growth rates; our linear growth/oscillation rates match local results in 
appropriate limits.

Appendix \ref{app:linear} gives the local and 
global forms of the linearized equations (Eqs. \ref{eq:eq4}-\ref{eq:eq8}) after separating dependent variables in the three directions.
We solve the global eigenvalue problem numerically by employing a pseudospectral method (\citealt{boyd2000}). The coupled first-order linear equations are 
converted into a matrix eigenproblem.  We de-dimensionalize the problem by scaling density and pressure with their values 
$(p_0[r_{\rm in}],~\rho_0[r_{\rm in}])$  at the inner 
radius $r_{\rm in}=0.1$ kpc; velocities are scaled by $c_0 \equiv (p_0[r_{\rm in}]/\rho_0[r_{\rm in}])^{1/2}$. All lengths are 
scaled by the outer radius $r_{\rm out}=100$ kpc. We find 
that the numerical eigen-solver is more robust if we de-dimensionalize the equations. Appendix \ref{app:pseudo} discusses the pseudospectral method in detail.

{\it Boundary conditions:}  The four first-order equations (Eqs. \ref{eq:ee1}-\ref{eq:ee4} for local analysis or \ref{eq:eq9}-\ref{eq:eq13} for global analysis in 
the transverse direction) require as many boundary conditions in general. A quick glance at Eqs. \ref{eq:ee1}-\ref{eq:ee4} suggests that the eigenvalues are
degenerate for global spherical modes with different $m$. The radial domain 
$[r_{\rm in},~r_{\rm out}]$ is mapped to $-1 \leq \xi \leq1$, in order to apply the pseudospectral method (see Appendix \ref{app:pseudo}). 
We assume no penetration through the outer radial direction ($v_{r 1} = 0$ at $r=r_{\rm out}$ or $\xi=1$). The other boundary conditions are
$v_{\theta 1} = 0$ at $\xi= 1$, $dv_{\theta 1}/d\xi = 0$ at $\xi=-1$ ($r=r_{\rm in}$), and $s_1=0$ at $\xi=-1$. We get similar results for other 
reasonable boundary conditions. The pseudospectral method is known to be extremely sensitive 
to the boundary conditions. Therefore, we chose our boundary conditions only after ensuring the convergence of eigenvalues and eigenmodes.

\subsection{Eigenvalues and numerical convergence}

\begin{figure}
 \includegraphics[width=.48\textwidth]{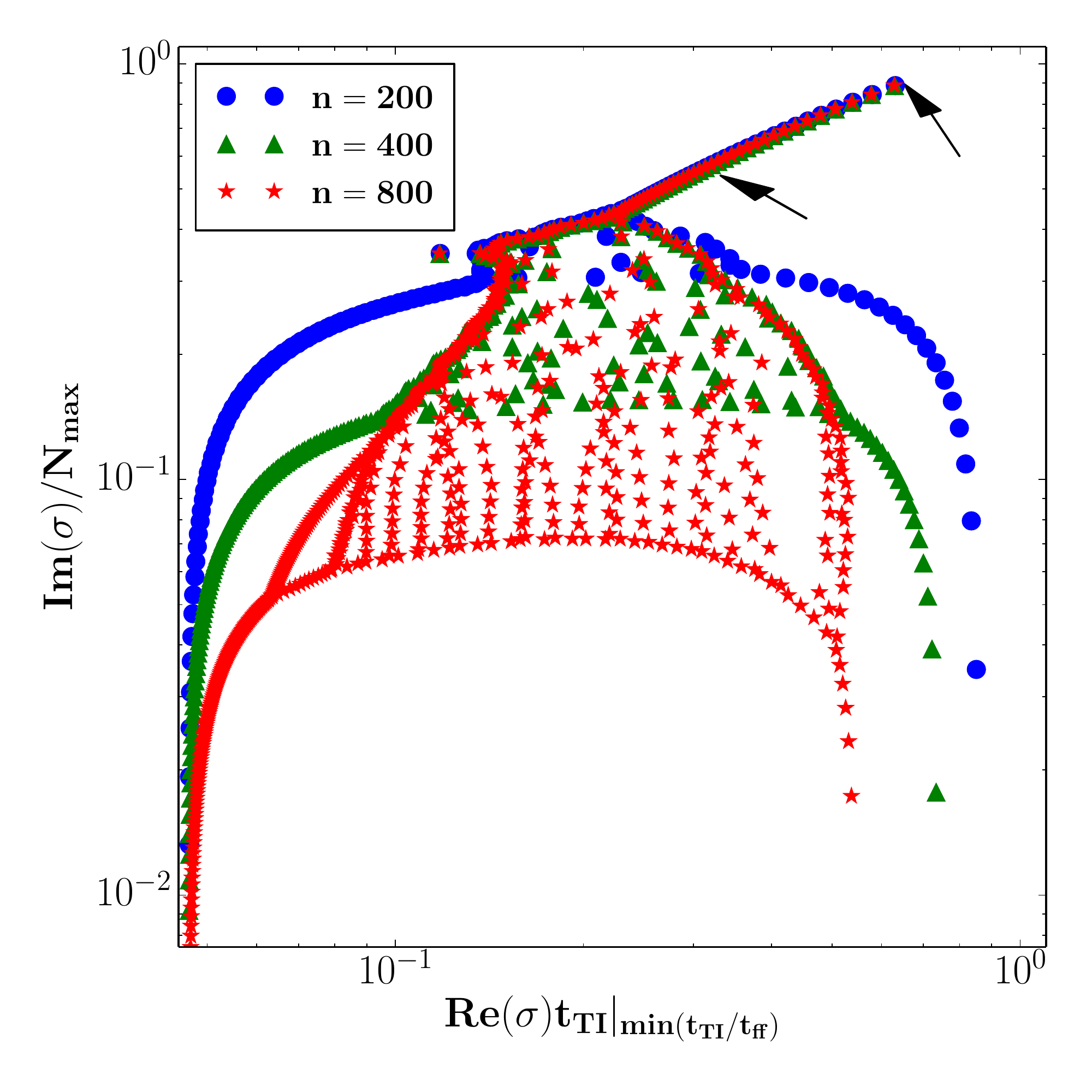}
 \caption{Convergence of eigenvalues of Eqs. \ref{eq:eq9} to \ref{eq:eq13}, for different resolutions ($n=200,~400,~800$) for our fiducial ICM confined by 
 NFW gravity. The real part of the eigenvalue is normalized with $N_{\rm max}$ and the imaginary part with $t_{\rm TI}$ evaluated at 
 min($t_{\rm TI}/t_{\rm ff}$). For these runs, $k_x r_{\rm out} = 100$. The arrows mark the eigenvalues for which the eigenmodes are shown in Figure 
 \ref{fig:fig04}. The fastest growing mode grows slightly slowly compared to the local growth rate at min($t_{\rm TI}/t_{\rm ff})$.}\label{fig:fig02} 
\end{figure}

Figure \ref{fig:fig02} shows numerical eigenvalues of the overstable (with a positive real $\sigma$) internal gravity modes, destabilized by thermal instability. 
More modes converge, 
and the eigenvalues fall on top of each other, as we increase the resolution. We only show the unstable modes; the sound wave is also affected by cooling and 
heating, but is damped (with a negative real part of $\sigma$) as the modes are effectively isentropic (\citealt{field65}). The physical modes show convergence with an 
increasing $n$ (number of Chebyschev polynomials or the G-L grid points in the eigenfunction expansion; see Eq. \ref{eq:exp}). 
The other modes are either physically spurious or unresolved. We only focus on the resolved modes. To 
compare  with local results, in Figure \ref{fig:fig02} the growth rate (real part of $\sigma$) is scaled with $t_{\rm TI}$ at min$(t_{\rm TI}/t_{\rm ff})$ 
in the equilibrium profile and the oscillation frequency (imaginary part of $\sigma$) is scaled with the maximum value of the 
BV frequency.

\begin{figure}
 \includegraphics[width=.48\textwidth]{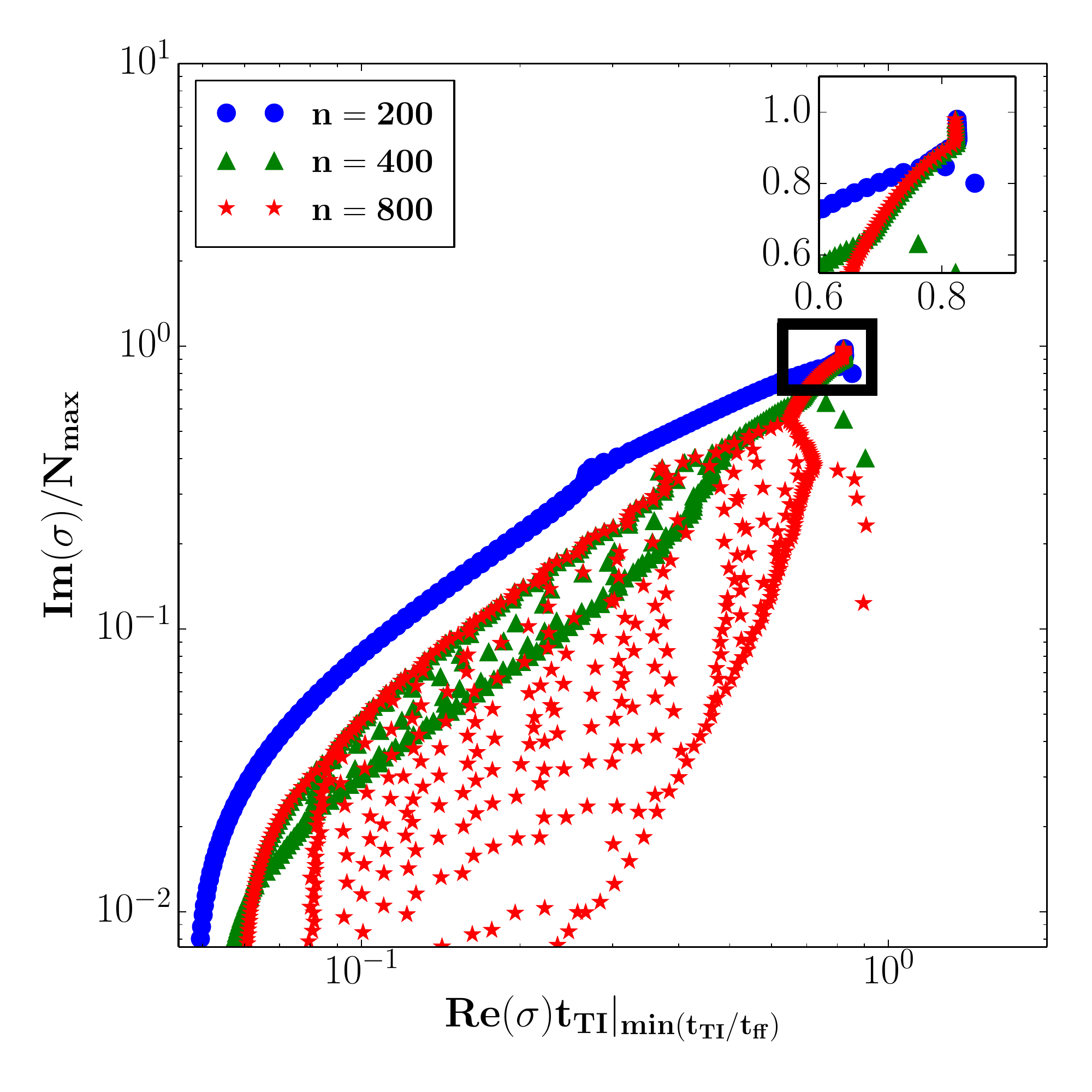}
 \caption{Convergence of eigenvalues for global linear analysis (Eqs. \ref{eq:ee1} to \ref{eq:ee4}) of thermal instability, using different resolutions 
for our fiducial ICM confined by NFW gravity. The degree of spherical harmonic mode is $l=10$ (corresponding to 
 $k_x (r_{\rm out}) r_{\rm out} \approx 10$). The inset shows the zoom-in on the converged modes. The growth rates are not substantially different from
 Figure \ref{fig:fig02}.}\label{fig:fig03} 
\end{figure}

Figure \ref{fig:fig03} shows the overstable internal gravity modes in a fully-global analysis in spherical geometry (taking a spherical harmonic basis function; Eqs. 
\ref{eq:ee1}-\ref{eq:ee4}). The value of $l$ is such that the transverse wavenumber is similar to the local analysis (Eqs. \ref{eq:eq9}-\ref{eq:eq13}) shown in Figure 
\ref{fig:fig02}. The growth rates of the fastest growing overstable modes in the global analysis are only slightly higher than the local analysis. Therefore, we conclude
that there is only a minor difference between the growth rates in global and local analyses. This is also verified for the truly global $l=2$ mode (see 
Table \ref{table:linanalysis}).

\subsection{Local and global gravity modes}

Assuming that the modes vary as $e^{(\sigma t+i\bm{k}\cdot \bm{x})}$ ($\bm{k} = \bm{\hat{x}} k_x + \bm{\hat{z}} k_z$ is the wavenumber), the local ($|k| H \gg 1$, 
$H$ is the entropy/pressure scale height) dispersion relation (using Eqs. \ref{eq:eq9}-\ref{eq:eq13}) is given as
\ba
\nonumber
&& \sigma^4 + \frac{\sigma^3}{t_{\rm cool}}\frac{d\ln \Lambda}{d\ln T}- \sigma^2 \left(ik_r g(r)   - {c_s}^2k^2 \right) \\
\nonumber
&&- \sigma\left( \frac{ik_rg(r)}{t_{\rm cool}}\frac{d\ln \Lambda}{d\ln T} + \frac{{c_s}^2k^2}{t_{\rm TI}}\right) + N^2 k_x^2 c_s^2  = 0.
\ea
Assuming, $N, t_{\rm TI}^{-1} \ll kc_s$ ($c_s \equiv [\gamma p_0/\rho_0]^{1/2}$ is the local sound speed) the low frequency g-modes have a dispersion relation,
$$
\sigma^2 -\frac{\sigma}{t_{\rm TI}} + \frac{{k_x}^2}{k^2} N^2  = 0,
\nonumber
$$
with the two roots being
$$
\sigma = \frac{1}{2t_{\rm TI}} \pm \frac{1}{2}\left( \frac{1}{t^2_{\rm TI}} - \frac{4 N^2 k^2_x}{k^2} \right)^{\frac{1}{2}}.
$$
For $k_x=0$, the $\sigma=0$ solution is trivial (i.e., $\rho_1,~v_{r1},~v_{\theta 1},~p_1=0$) in the local ($|k| H \gg 1$) limit. The $\sigma = t_{\rm TI}^{-1}$ solution 
does not exist  
if heating balances cooling in shells ($q^+ = \langle q^- \rangle $ or $q^+_1 = q^-_1$, as in our set up), but exists 
as the local isobaric thermal instability if 
thermal balance does not hold for {\it perturbations} (e.g., if $q^+_1=0$, or in other words, if thermal balance is only imposed in equilibrium 
not on perturbations).\footnote{\citet{balbus88} has argued that there is no $k_x=0$ (purely radial) 
isobaric thermally instability mode. However, linear analysis (Eqs. \ref{eq:eq9}-\ref{eq:eq13}) shows, as long as heating and cooling do not exactly balance in 
shells (i.e., $q^+_1 \ne q^-_1$), that it will exist as a robust instability with purely growing modes with $\sigma=t_{\rm TI}^{-1}$ and $p_1/p_0 \ll \rho_1/\rho_0$.} 
%{\bf{Note that in our system, heating is equal to shell-averaged cooling exactly like in our simulations. However, for $k_x = 0$, this prescription imposes
%exact thermal equilibrium on perturbations as well. Our motivation is to look for modes with $k_x\neq0$ to match with
%the results of simulations and in that case, only the background medium is in exact thermal balance(which also justifies our Eulerian approach).}}\footnote{\citet{balbus88} has argued that there is no $k_x=0$ (purely radial) 
%isobaric thermally instability mode. However, linear analysis (Eqs. \ref{eq:eq9}-\ref{eq:eq13}) shows, as long as heating and cooling do not exactly balance in 
%shells (i.e., $q^+_1 \ne q^-_1$), that it will exist as a robust instability with purely growing modes with $\sigma=t_{\rm TI}^{-1}$ and $p_1/p_0 \ll \rho_1/\rho_0$.} 

\begin{figure}
 \includegraphics[width=.5\textwidth]{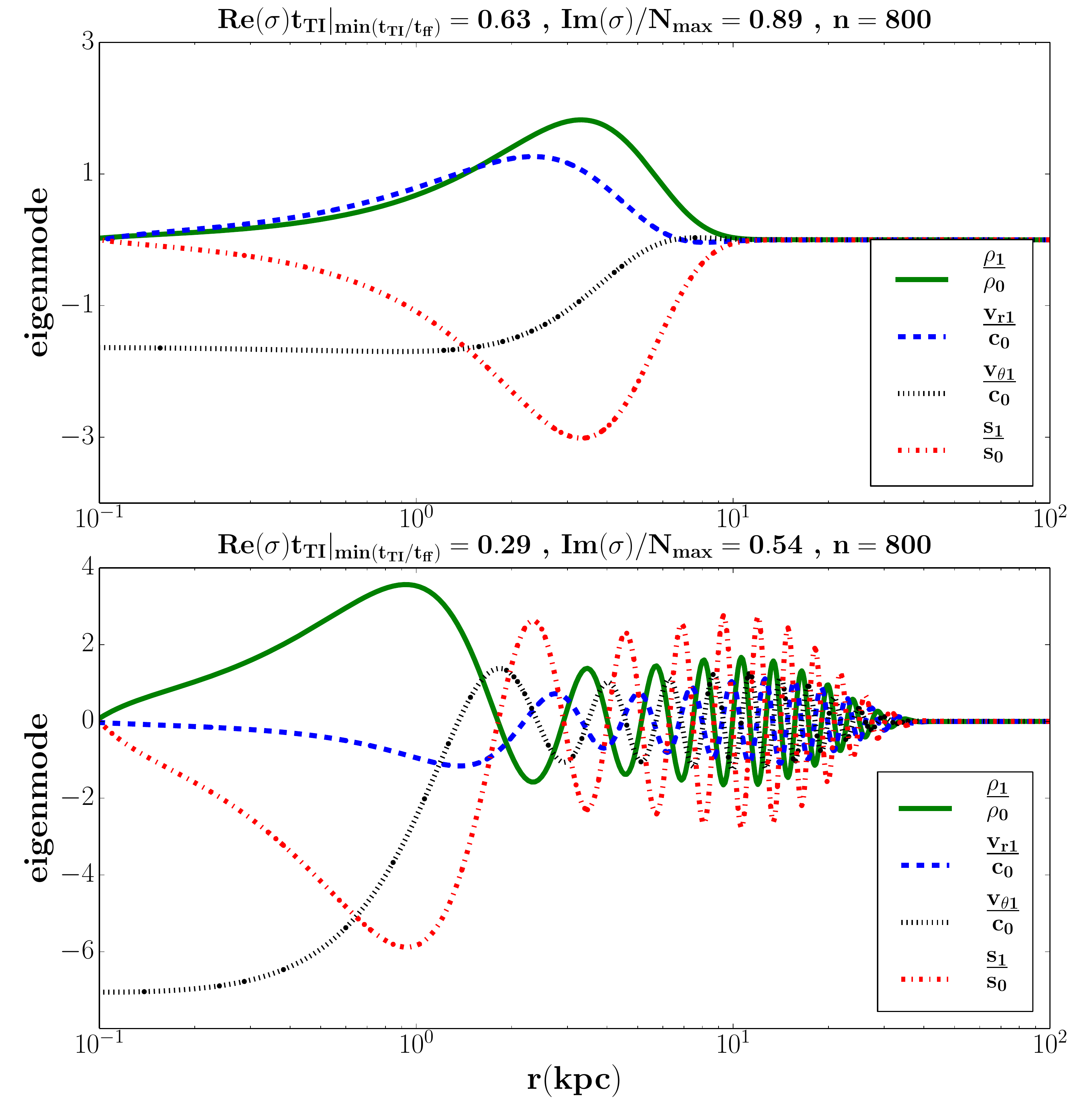}
 \caption{Eigenmodes for two different thermally unstable gravity modes with different eigenvalues, as indicated by arrows in Figure \ref{fig:fig02}. 
 We have verified that we obtain the same eigenmodes for different resolutions that show convergence for eigenvalues 
 ($n=200,~400,~800$ in Fig. \ref{fig:fig02}).}\label{fig:fig04} 
\end{figure}

\begin{figure}
 \includegraphics[width=.5\textwidth]{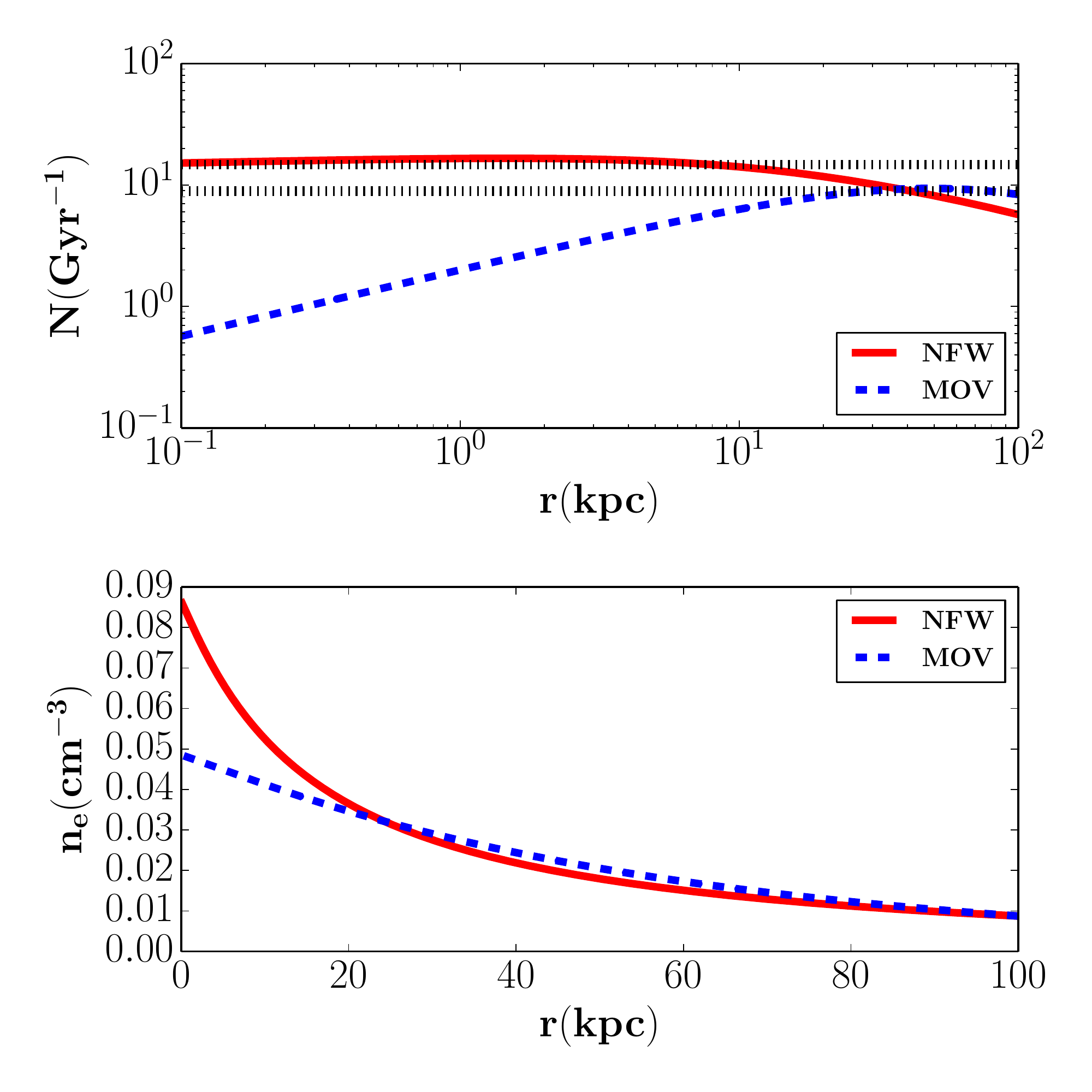}
 \caption{{\it Top panel:} The local BV frequency as a function of radius for NFW ($K_0=8$ keV cm$^2$) and MOV ($K_0=23$ keV cm$^2$) gravities. The horizontal
 dotted lines represent the imaginary part of the growth rate ($\sigma_i/N_{\rm max}=0.89,~0.54$) for the modes shown in Figure 
 \ref{fig:fig04}. Eigenfunctions are confined to radii where $\sigma_i < N$ for both modes, as expected. Similarly, as expected from the shape of $N$ 
versus $r$ for MOV gravity, we verify that the lowest radial wavenumber modes are confined to $r \approx 60$ kpc (where $N$ peaks). The 
 BV frequency has a different profile for NFW and MOV gravities as it is proportional to the local gravitational acceleration, which is qualitatively different for the 
 two cases (Figure \ref{fig:fig01}). {\it Bottom panel:} The electron number density as a function of radius for NFW and MOV gravities.}
 \label{fig:fig05} 
\end{figure}

Since the slowest varying mode in the $r-$ direction (i.e., with lowest $k_r$) are the easiest to resolve, and they have the largest oscillation frequency of gravity 
modes ($\sigma \approx \pm i k_x N/[k_x^2 + k_r^2]^{1/2}$), the modes with the highest imaginary values of $\sigma$ are resolved most easily 
(see Fig. \ref{fig:fig02}). 
The profile of BV frequency (Eq. \ref{eq:BV}) for cool-core clusters is peaked toward the
center and fall off at large radii (solid red line in Fig. \ref{fig:fig05}). Therefore, the internal gravity modes become evanescent at large radii where 
$| \sigma_i| > N$, and the eigenmodes are confined
to small radii in the center (e.g., \citealt{BalHaw1995}). Gravity modes with a higher number of nodes and antinodes (i.e., higher $k_r$) have a smaller oscillation 
frequency, and hence are evanescent at larger radii. The two gravity eigenmodes shown in Figure \ref{fig:fig04} clearly illustrate this; a higher $k_r$ 
corresponds to a smaller oscillation frequency and a more extended radial structure. As shown in Figure \ref{fig:fig05}, the local BV frequency is peaked at larger 
radii for MOV gravity, therefore the resolved modes are confined at large radii (we have verified this) where the cooling time is long, and hence the growth rate is 
smaller (see Table \ref{table:linanalysis} and red squares in Fig. \ref{fig:fig06}).

\subsection{Growth rates in plane-parallel and spherical atmospheres}
\label{sec:lin_sphvscart}

\begin{figure}
 \includegraphics[width=.5\textwidth]{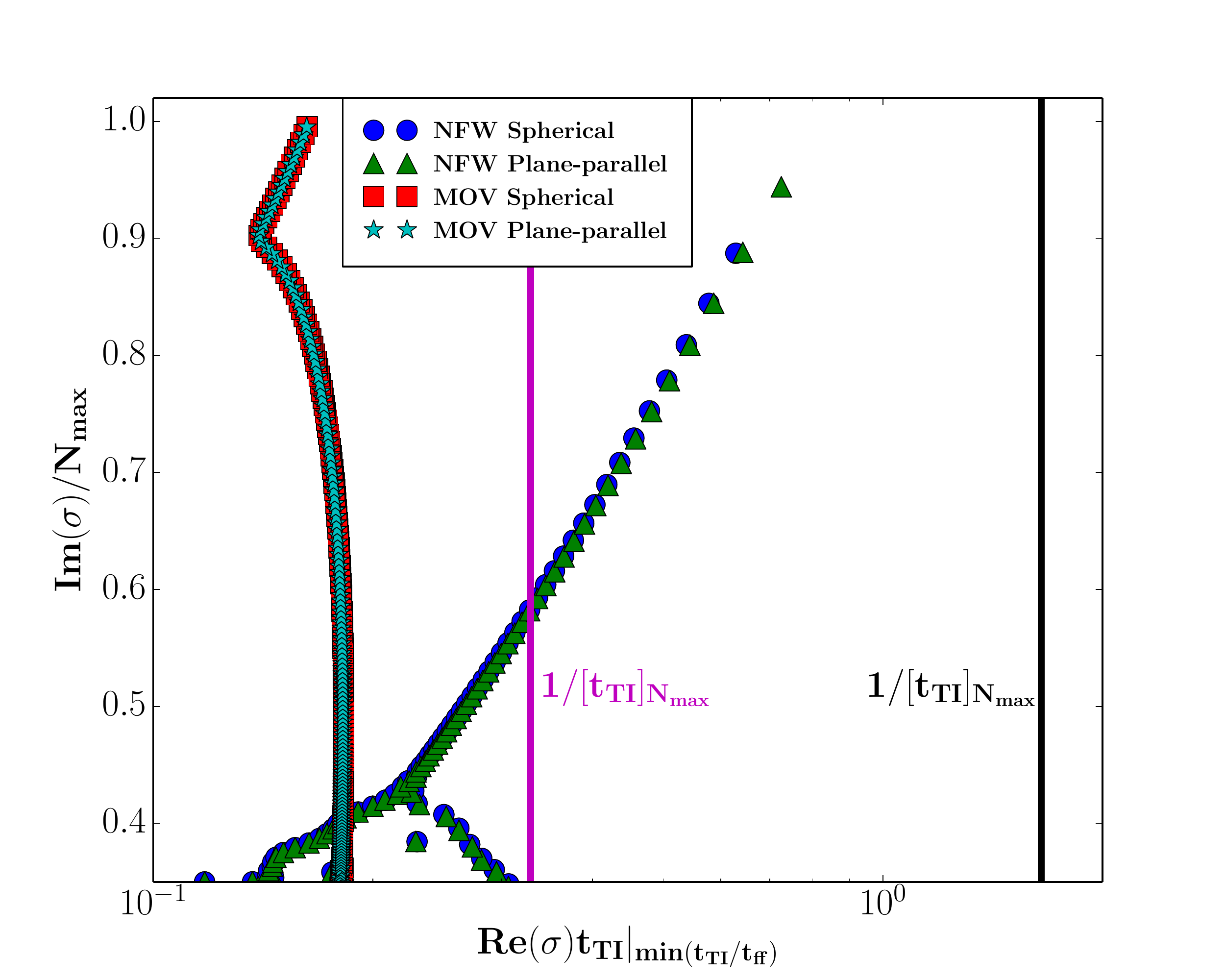}
 \caption{The converged thermally unstable internal gravity modes for spherical and plane-parallel atmospheres for NFW and MOV gravities.The only difference 
 between spherical and plane-parallel gravities is that $(1/r^2) \partial/\partial r (r^2 \rho_0 v_{r1})$ term in Eq. \ref{eq:eq9} is replaced by $\partial/\partial 
 z(\rho_0 v_{z1})$. The growth rates for converged modes are almost equal (for MOV they coincide), suggesting that there is no effect of geometry 
 on the linear growth rate. 
 Here $n=800$ and $k_x r_{\rm out} = 100$. The vertical lines correspond to the local thermal instability growth rates ($t_{\rm TI}$) measured at 
 the peak of the local BV frequency ($N_{\rm max}$). The magenta lines correspond to MOV gravity and black lines 
 to NFW gravity. Note that the $N_{\rm max}$ peak for NFW gravity occurs close to the center where density is high, 
 but for MOV gravity $N_{\rm max}$ it is further out where the density is much lower (Fig. \ref{fig:fig05}).}\label{fig:fig06} 
\end{figure}

Figure \ref{fig:fig06} compares the converged growth rates of thermally unstable internal gravity waves in spherical and plane parallel atmospheres for NFW and 
MOV gravities. It shows that there is no effect of geometry on the growth rate, unlike expected (e.g., \citealt{ashmeet14}). Also note that the difference between 
spherical and plane-parallel atmospheres is slightly more for NFW gravity because the converged modes are centrally concentrated around $r=0$ and truly global, 
whereas for MOV gravity the small radial wavenumber modes are confined over $\Delta r/r \lesssim 1$ close to $N_{\rm max}$ and are essentially local.

\section{Numerical simulations}
\label{sec:num_sims}
Now that global stability analysis clearly shows that there are only minor differences in thermal instability growth rates  
in plane-parallel and spherical atmospheres, we turn to idealized 2-D/axisymmetric nonlinear numerical simulations to compare the two. 

We perform 2-D axisymmetric simulations with both Cartesian (also referred to as `plane-parallel') and spherical atmospheres (setup is described in section 
\ref{sec:sim_setup}). While NFW gravity simulations
are carried out in both plane-parallel and spherical geometries, simulations using MOV and MSQP gravities are only carried out in Cartesian geometries. A
 comparison of spherical and plane-parallel NFW simulations clearly shows that there is very little influence of geometry on the onset of
  cold gas condensation. In both 
 geometries, in situ condensation requires a smaller min($t_{\rm cool}/t_{\rm ff}$) compared to condensation anywhere in the simulation box (section \ref{sec:NFW_runs}). The Cartesian NFW, MOV and MSQP simulations are used to test the influence of gravitational acceleration and density profiles 
 (Eqs. \ref{eq:g_NFW}- \ref{eq:g_MSQP}) on cold gas condensation (section \ref{sec:diff_g}).

\subsection{Simulation setup}
\label{sec:sim_setup}
We use the {\tt ZEUS-MP} code (\citealt{hayes06}) to solve Euler equations with heating, cooling, and gravity (Eqs. \ref{eq:eq1}-\ref{eq:eq3}). The initial condition
consists of a hydrostatic equilibrium profile (described in section \ref{sec:sec2}), superposed with isobaric small density perturbations. We also carry out 
simulations using different potential wells (NFW, MOV, MSQP; Eqs. \ref{eq:g_NFW}-\ref{eq:g_MSQP}) and different geometries (spherical \& plane-parallel).

\begin{figure}
 \includegraphics[width=.5\textwidth]{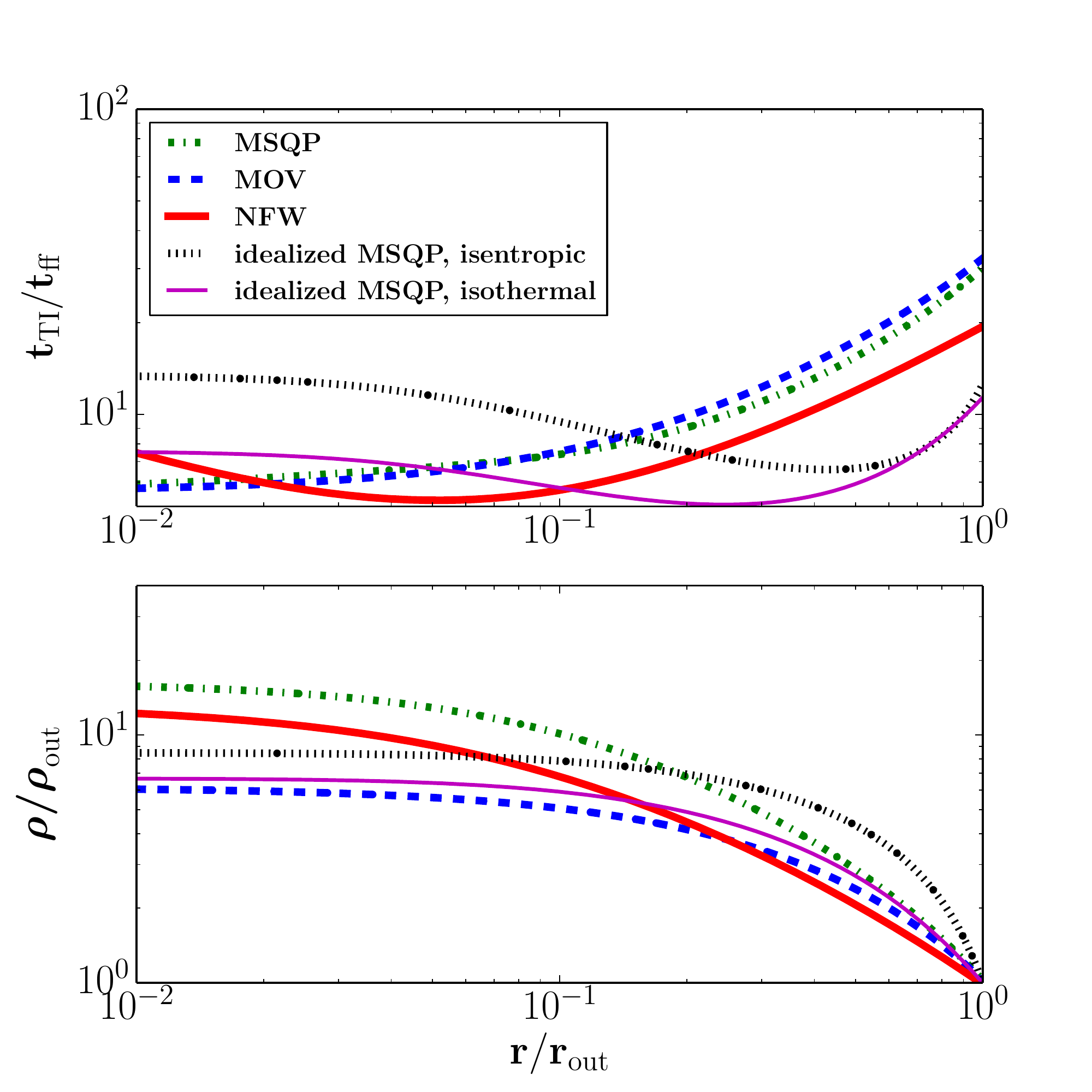}
 \caption{{\it Top panel:} The profiles of the ratio of the local thermal instability timescale and the local free-fall time ($t_{\rm TI}/t_{\rm ff}$) 
 for various equilibrium atmospheres
 with min($t_{\rm TI}/t_{\rm ff} \approx 5$): NFW, MOV, MSQP, and idealized MSQP (isothermal and isentropic). {\it Bottom panel:} 
 The normalized gas density profiles for the same 
 runs. While MSQP atmosphere uses Eq. \ref{eq:ent} for the entropy profile, the  idealized MSQP profiles are isothermal/isentropic and  identical to Eq. 
 9/10 of \citet{mccourt12}. Also, while NFW and idealized MSQP profiles have a local minimum, the MOV and MSQP profiles have a minimum at the center. 
 }
 \label{fig:fig07}
\end{figure}

In addition to the runs with cluster-like parameters, we perform a few runs with the idealized setup similar to \cite{mccourt12} (isothermal/isentropic 
corresponding to their Eqs. 9/10; these also use a $T^{1/2}$ cooling function cut off at 1/20). Table \ref{table:sims} 
lists all our runs. Figure \ref{fig:fig07} shows the profiles of the ratio of the thermal instability timescale and the free fall time ($t_{\rm TI}/t_{\rm ff}$) in
the unperturbed state of some of our model atmospheres (NFW, MOV, MSQP, and idealized MSQP). This ratio is important to interpret cold gas 
condensation in our setup.

\subsubsection{Grids and geometries}

The Cartesian runs have a resolution of $512\times256~(N_z \times N_x)$,  while the spherical runs have a resolution of 
$256\times256~(N_r \times N_\theta)$. We have a comparable spatial resolution for the two geometries in the
direction along gravity. The Cartesian grid runs from $-100$ kpc to $100$ kpc in all  three dimensions, while the spherical grid has the following range 
in the three directions: 1 kpc $\leq r \leq$ 100 kpc, $0 \leq \theta \leq \pi$ and $0 \leq \phi \leq 2\pi$. 

The radial grid in spherical runs is logarithmic, with equal number of grid points from 1 to 10 kpc and 10 to 100 kpc. The Cartesian $z$-grid is partly 
logarithmic ($-100\ \rm  kpc$ to $-1\ \rm  kpc$, $1\ \rm  kpc$ to $100\ \rm  kpc$) and partly uniform($-1\ \rm  kpc$ to $1\ \rm  kpc$), to match the resolution
of the spherical simulations in the direction of gravity. Cartesian runs corresponding to MOV and MSQP gravities have a similar grid resolution.  The idealized 
MSQP runs use a uniform grid throughout.

\begin{table*}
\caption{List of idealized nonlinear simulations \& their parameters}
\centering
\begin{tabular}{c c c c c c c}
\hline\hline
Gravity & $K_{\rm 0}$ & ${\rm min}(t_{\rm TI}/t_{\rm ff})$ & total $M_{\rm cold}/M_{\rm tot}^a$ & $ M_{\rm cold}/M_{\rm tot}^b$ & $r_{\rm cold, max}$ & runs till \\
 & (keV cm$^2$) & (in &   &  at ${\rm min}(t_{\rm TI}/t_{\rm ff})$  & (kpc) & (Gyr) \\
  & & equilibrium) & [anywhere]  &   [in situ]  & &  \\
\hline
spherical, NFW &  $ 2.0$ & $2.6$ & $.0016$ & $0.12$ &$24.4$ & $15$ \\
               &  $ 5.0$ & $5.2$ & $.0014$ & $0.002$ &$11.4$ & $15$ \\
               &  $ 8.0^\ddag$ & $7.3$ & $3.5 \times 10^{-4}$ & $2.1 \times 10^{-4}$ &$15.5$ & $15$ \\
               & $ 10.0$ & $8.6$ & $2.2 \times 10^{-4}$ & $0.0$ &$6.6$ & $15$ \\
               & $ 20.0$ & $13.6$ & $0.0$ & 0.0 & - & $15$ \\
               & $ 30.0$ & $17.6$&$0.0$&$0.0$&-&$15$\\
               & $ 45.0$ & $22.5$&$0.0$&$0.0$&-&$15$\\
\hline
Cartesian, NFW &  $ 2.0$ & $2.6$ & $0.09$ & $0.123$ &$13.5$ & $15$ \\
               &  $ 5.0$ & $5.2$ & $0.07$ & $0.005$ &$19.3$ & $15$ \\
               &&&$(0.08)$ & $(0.006)$ & $(19.3)$&$(10.9)$\\
               &  $ 8.0{^\ddag}$ & $7.3$ & $0.08$ & $0.003$ &$9.9$ & $15$ \\
               &&&$(0.09)$ & $(0.004)$ & $(9.9)$&$(10.9)$\\
               &&&$(.12)$&$(0.005)$&$(9.9)$&$(6.5)$\\
               & $ 10.0$ & $8.6$ & $0.04$ & $0.0$ &$19.3$ & $15$ \\
               &&&$(0.05)$ & $(0.0)$ & $(19.3)$&$(10.9)$\\
               & $ 20.0$ & $13.6$ & $0.03$ & $10^{-5}$ &$17.68$ & $15$ \\
               &&&$(0.02)$ & $(10^{-5})$ & $(17.68)$&$(10.9)$\\
               & $ 30.0$ & $17.6$&$.02$&$0.0$&$16.53$&$15$\\
               &&&$(0.007)$ & $(0.0)$ & $(2.31)$&$(10.9)$\\
               & $ 45.0$ & $22.5$&$0.0$&$0.0$&-&$15$\\
\hline               
Cartesian, MSQP& $7.5$ & $5.3$ & $0.39$ & $0.64$ &$19.8$ & $6.5$ \\
               & $15.0$& $11.85$ & $0.4$ & $0.61$&$11.8$ & $6.5$ \\
               & $28.0$ & $19.2$ & $0.19$ & $0.33$ &$13.8$&$6.5$ \\
               & $42.0$& $25.8$ & $0.0$ & $0.0$ & $-$ &$6.5$ \\
%               & $60.0$& $33.3$& $0.0$ & $0.0$ &-& $6.5$ \\
\hline
Cartesian, MOV & $18.0$ & $5.4$ & $0.36$ &$0.61$ &$38.7$& $6.5$ \\
               & & &$(0.49)$&$(0.66)$&$(38.7)$&$(10.9)$ \\
  & $40.0$ & $12.8$ & $0.34$ & $0.65$ & $42.3$ & $10.9$ \\
  & $60.0$ &$19.6$ &$0.002$&$0.012$&$11.1$&$10.9$\\
  & $80.0$ &$26.2$ &$0.0$&$0.0$&-&$10.9$\\
  & &  &$(10^{-3})^{\dag}$  & $({1.3 \times 10^{-3}})^{\dag}$ & $(2.0)^{\dag}$& $(15)^{\dag}$\\
  & $100.0$ &$32.2$ &$0.0$&$0.0$&-&$10.9$\\
 (cooling/heating also in $ |z| \leq 1~{\rm kpc}$) & $80.0^{\dag}$ & $26.2^{\dag}$ &$0.18^{\dag}$ &$0.32^{\dag}$&$26.4$ $^{\dag}$&$15^{\dag}$\\
\hline\hline
 &  &   &  &  & $r_{\rm cold, max}$& runs till \\
  &  &   &  &  & (code units) & (code units) \\
\hline
Cartesian, idealized MSQP (isentropic)  & - & $0.11$ & $0.85$ & $0.14$ &$2.0$& $180$ \\
   (same as \citealt{mccourt12}; Eqs. 10a-b)  & - & $0.37$ & $0.66$ & $0.01$ &$2.0$& $180$ \\
         $z_{\rm in},~z_{\rm out}=-2,~2$                     & - & $1.1$ & $0.29$ & $2.4 \times 10^{-4}$ &$2.0$& $180$ \\
     $\Lambda = \Lambda_0 T^{1/2}$ and cuts-off to 0  below     & - & $3.33$ & $0.1$ & $0.0$ &$0.49$& $180$ \\
      0.05 times the initial midplane temperature;                         & - & $6.6$ & $0.004$ & $0.0$ &$0.53$& $180$ \\
      $ t_{\rm TI}/t_{\rm ff} $ is controlled by changing $\Lambda_0$                      & - & $11.0$& $0.0$ & $0.0$ &-& $180$ \\
\hline
Cartesian, idealized MSQP (isothermal)  & - & $0.094$ & $0.89$ & $0.28$ &$2.0$& $180$ \\
   (same as \citealt{mccourt12}; Eq. 9)  & - & $0.31$ & $0.78$ & $0.08$ &$2.0$& $180$ \\
         $z_{\rm in},~z_{\rm out}=-2,~2$                     & - & $0.94$ & $0.37$ & $0.003$ &$1.7$& $180$ \\
     $\Lambda = \Lambda_0 T^{1/2}$ and cuts-off to 0  below     & - & $2.8$ & $0.06$ & $0.0$ &$0.9$& $180$ \\
      0.05 times the initial midplane temperature;                         & - & $5.6$ & $0.01$ & $0.0$ &$0.2$& $180$ \\
      $ t_{\rm TI}/t_{\rm ff} $ is controlled by changing $\Lambda_0$                      & - & $9.4$& $1.2 \times 10^{-4}$ & $0.0$ &$0.2$& $180$ \\
                                                                                              & - & $13.1$& $0.0$ & $0.0$ &-& $180$ \\
\hline
\end{tabular}
\textbf{Notes:} $^a$ The ratio of time-averaged cold gas mass and total gas mass within $1 {\rm kpc} \leq r \leq r_{\rm cold, max}$. $r_{\rm cold, max}$ is the 
maximum distance from the center (at all times) at which cold gas is observed.  Some simulations are run for 6.5 and 10.9 Gyr because they take very long; 
for a uniform comparison the same run is sometimes analyzed till different times. \\
$^b$ The ratio of time-averaged cold gas mass and total gas mass within $0.9 H$ and $1.1 H$ ($H$ corresponds to min[$t_{\rm TI}/t_{\rm ff}$]). \\
$^\ddag$ The fiducial NFW runs.  \\
$^\dag$ MOV runs with and without heating and cooling at $|z| \leq 1$. 
\label{table:sims}
\end{table*}

\subsubsection{Initial and boundary conditions} 
\label{sec:sims}

For all runs we seed initial isobaric density perturbations such that the maximum $\delta \rho/ \rho_0 \approx 0.3$ (the runs focusing 
on the early linear stage of thermal instability,  shown in Fig. \ref{fig:fig10}, use a smaller maximum amplitude of 0.01), where
$\delta \rho = \rho -\rho_0$ and  $\rho_0$ is the equilibrium density as a function of $r$~(or $z$). 
The density fluctuations are isotropic and homogeneous (in the three dimensional Cartesian sense), given by
$$
\frac{\delta \rho}{\rho_0} = \sum_{k,l,m} a_{k,l,m} \cos \left ( \frac{2\pi (k x + l y + m z) }{r_{\rm out}} + \phi_{k,l,m} \right ),
$$
where $\phi_{k,l,m}$ is the uniformly distributed random phase, the random mode amplitudes $a_{k,l,m}$ have a power spectrum
$\propto (k^2+l^2+m^2)^{-1/2}$ for $2 \leq  (|k|,|l|,|m|) r_{\rm out}/2 \pi \leq 10$, where $r_{\rm out} = 100\ \rm  kpc$ is the outer radius. 
We choose the same mode amplitudes and phases for all spherical and plane-parallel runs, and therefore the density perturbations 
($\delta \rho/\rho_0$) are identical in all our runs. This helps us make a precise comparison between our different runs.

At outer boundary the electron number density is fixed to be $n_{e,{\rm out}} = 0.00875~{\rm cm}^{-3}$. For spherical simulations, the boundary conditions 
in the radial direction allow outflow at the inner boundary ($r_{\rm in}=1$ kpc) and inflow at the outer boundary ($r_{\rm out}=100$ kpc), with the density 
and internal energy density fixed to their equilibrium values at the outer boundary. The boundary conditions for $\theta$ and $\phi$ directions are, 
respectively, reflective/axisymmetric and periodic. For Cartesian runs we apply inflow boundary conditions at the two vertical boundaries 
(at $z=\pm 100$ kpc) 
and we fix the density and internal energy densities to their equilibrium values. We apply reflective boundary condition in $x$ and $y$ directions.

\begin{figure*}
 \includegraphics[width=.7\textwidth]{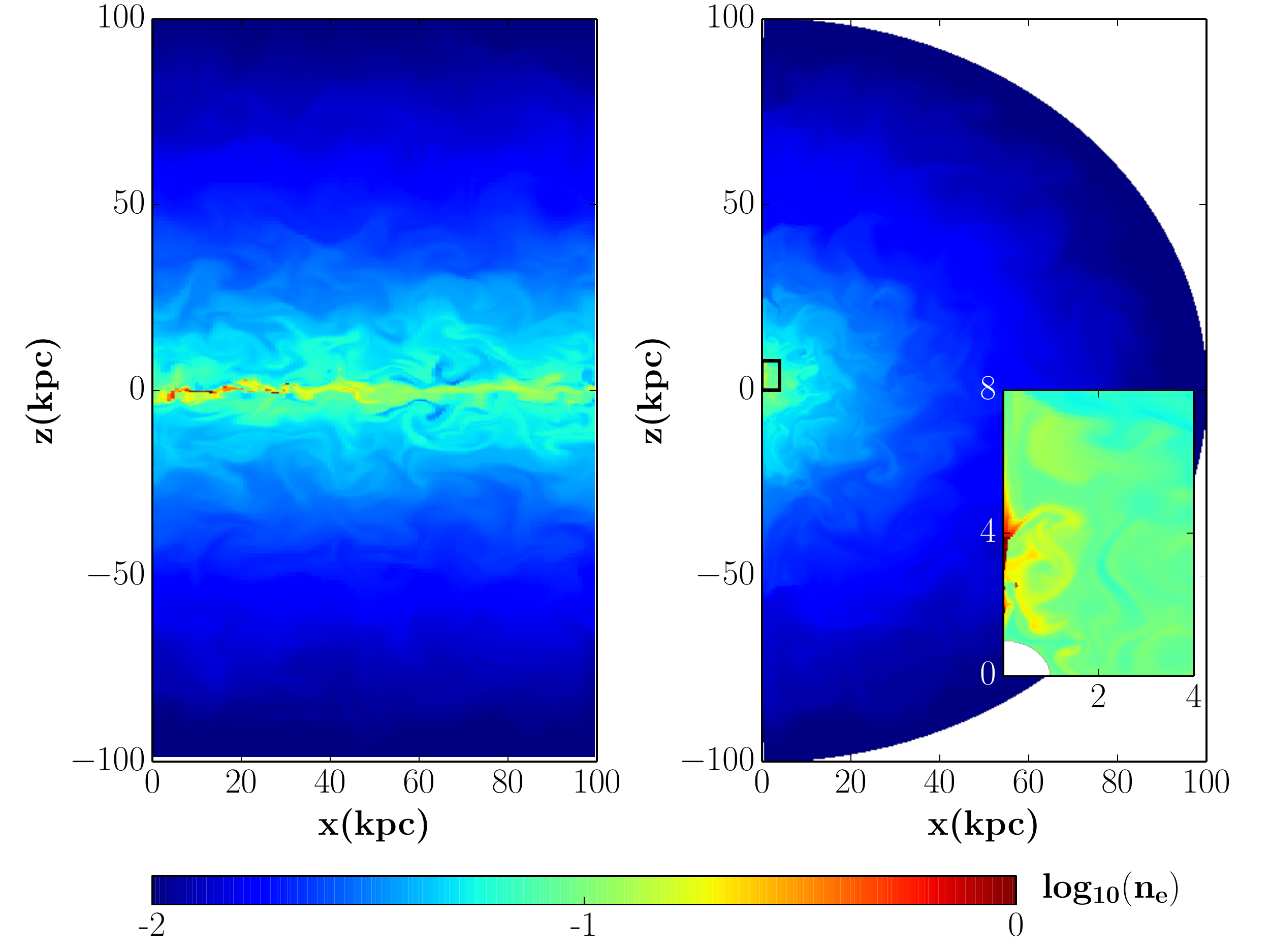}
 \caption{The electron number density snapshots at 1 Gyr (just when cold gas starts to condense) for NFW runs with ${\rm min} (t_{\rm TI}/t_{\rm ff}) \approx 7.3$ 
 in the equilibrium (unperturbed) profiles. {\it Left panel:} plane-parallel atmosphere (only half of the horizontal extent is shown); 
 {\it Right panel:} spherical atmosphere. The inset in the right panel 
 shows the zoomed-in view to highlight the cold gas in the spherical simulation. 
 A key difference between the two setups is that the mass of the cooler gas, out of which cold gas condenses, is much smaller in the 
 spherical atmosphere because of geometry.}
 \label{fig:fig08}
\end{figure*}

 \begin{figure*}
 \includegraphics[width=.95\textwidth]{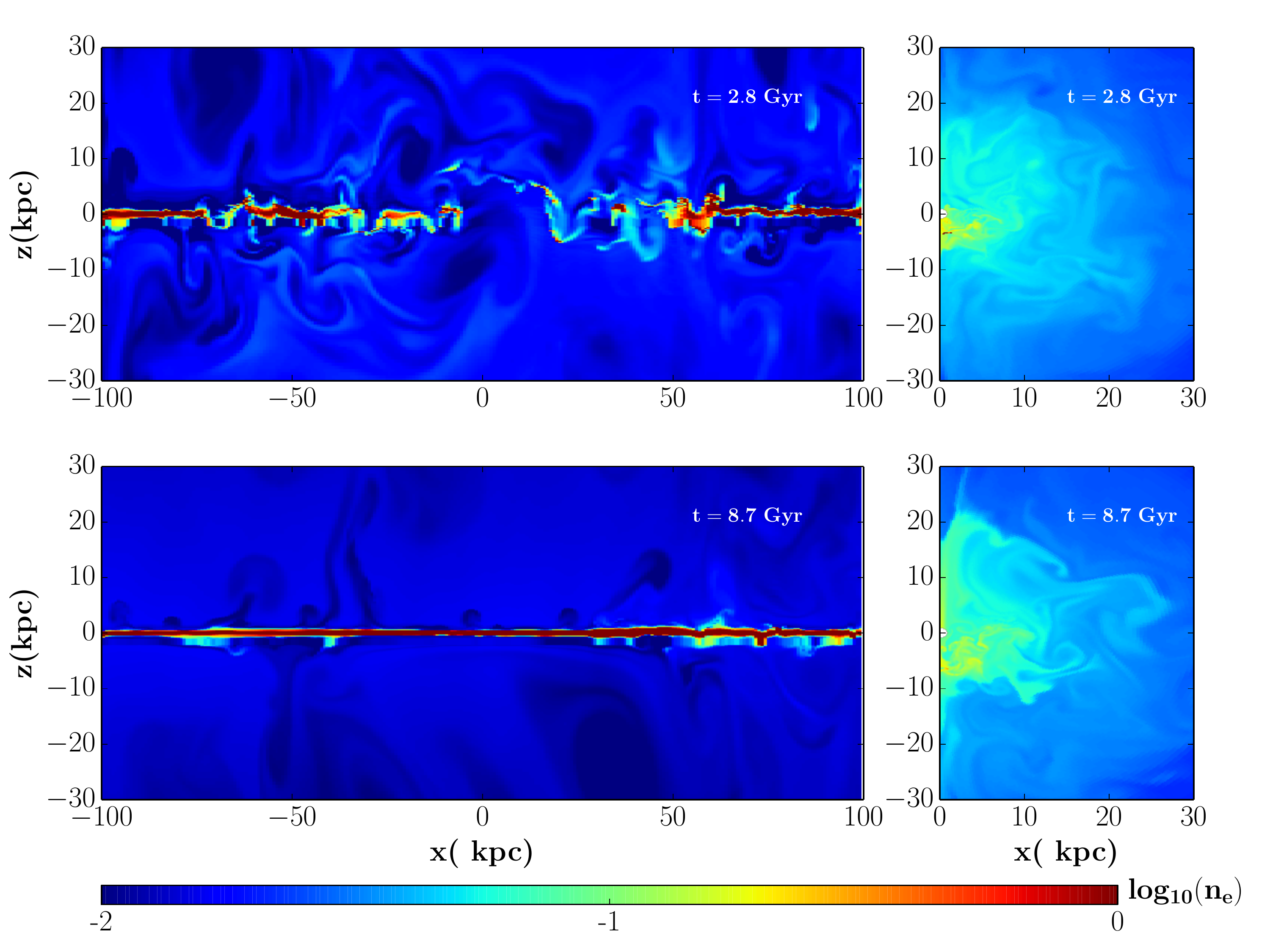}
 \caption{Snapshots of electron number density at two times for plane-parallel ({\it left}) and spherical ({\it right}) fiducial NFW runs; 
 early times when there is extended cold gas in the plane-parallel setup and late times when most 
 cold gas is in the midplane. A key difference between the two setups is that the center is excised in the spherical runs, but cold gas has nowhere to go 
 except accumulate at the center in Cartesian runs. Also, the hot gas density is much lower for Cartesian setup 
 as compared to spherical.}\label{fig:fig09}
\end{figure*}
  
\subsection{Influence of geometry: results from NFW runs}
\label{sec:NFW_runs}

In this section we compare the results from spherical and plane-parallel NFW simulations to assess the role of geometry on cold gas condensation. As with global 
linear analysis in section \ref{sec:lin_sphvscart}, numerical simulations show only minor differences between spherical and Cartesian runs.

Figure \ref{fig:fig08} shows the electron number density snapshots at 1 Gyr, just at the beginning of cold gas condensation in our fiducial NFW simulations 
($K_0=8$ keV cm$^2$, min$(t_{\rm TI}/t_{\rm ff})=7.3$). The key 
difference between the plane-parallel and spherical atmospheres is that the former has a much larger volume of dense gas out of which cold gas can condense 
(the volume prone to condensation in Cartesian setup is $2 r_{\rm cold, max}[100~\rm kpc]^2$, much larger than in spherical geometry 
$[4\pi/3] r_{\rm cold, max}^3$; $r_{\rm cold, max}$ is the radius/height within which condensation occurs).
This results in a much larger amount of cold gas condensing in plane-parallel atmospheres. While cold gas is easily seen in the midplane of the Cartesian setup, it
is only seen in the zoomed-in inset for the spherical setup.

Figure \ref{fig:fig09} shows the zoomed-in electron density snapshots for the Cartesian and spherical runs at times when there is significant cold gas 
present in the midplane of the plane-parallel run. One 
can see infalling cold/dense gas condensing out of the hot atmosphere and rising low density plumes; this is very prominent in the Cartesian run. 
Extended condensation 
is suppressed later because the dense hot gas has been depleted and has accumulated at the center. Depletion of hot gas density is much more prominent in
the Cartesian setup as large amount of cold gas condenses out and our ansatz of heating balancing cooling overheats the upper layers. 
Condensation is much smaller in spherical geometry because the volume of gas susceptible to 
cooling is smaller and there is a huge mass reservoir available to compensate for mass dropout in the core. While the amount of cold gas condensing out strongly
depends on geometry, the min$(t_{\rm TI}/t_{\rm ff})$ threshold for both in situ and anywhere condensation is insensitive to it.

\subsubsection{Growth rate of perturbations from simulations} 

\begin{figure}
 \includegraphics[width=.5\textwidth]{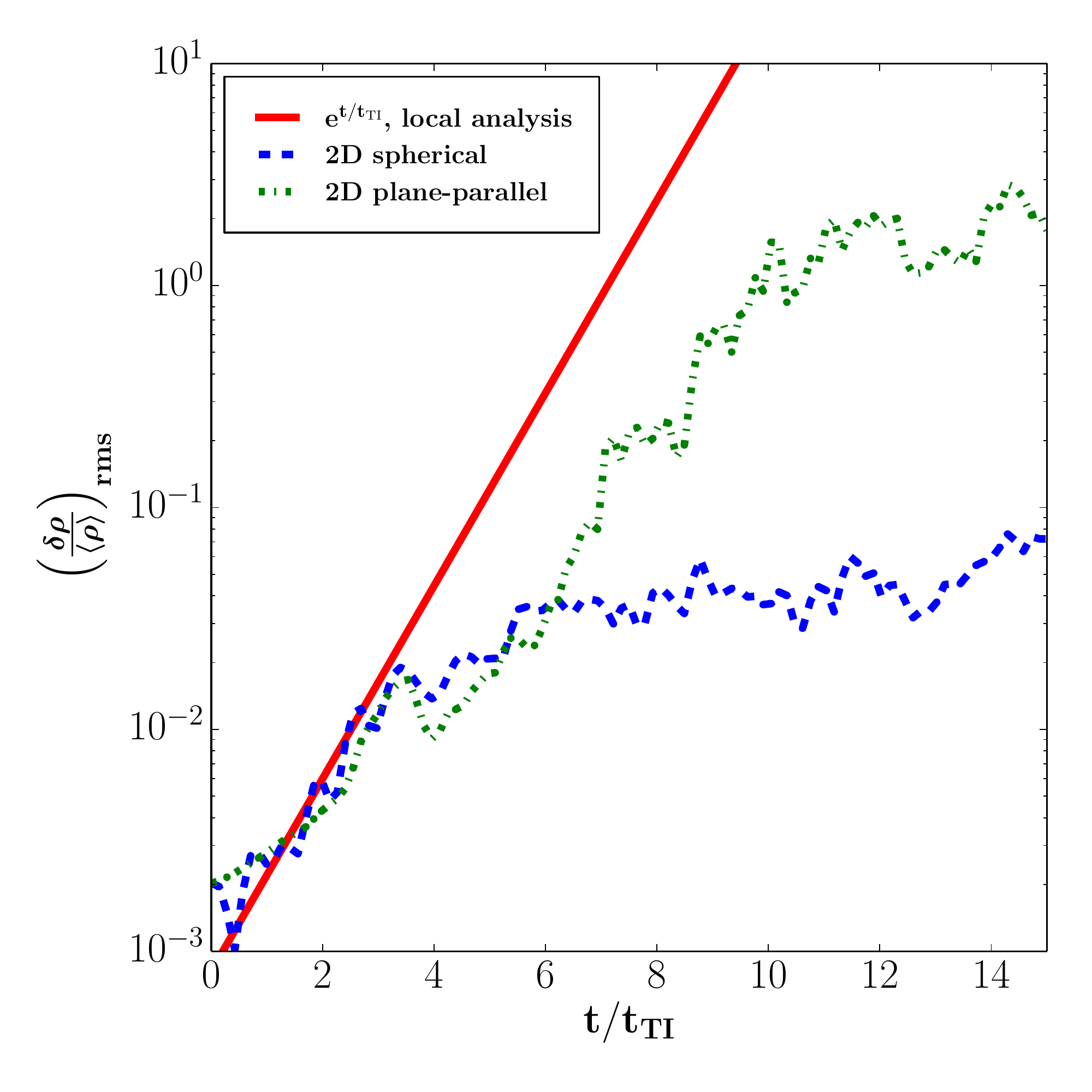}
 \caption{The amplitude of rms density fluctuations $(\delta \rho /\langle \rho \rangle)_{\rm rms}$, where $\delta \rho = \rho - \langle \rho \rangle$ and 
 $\langle \rangle$ represents shell averaging) measured between $0.9-1.1 H$ ($H$ is the radius/height corresponding to min[$t_{\rm TI}/t_{\rm ff}$] in the 
 unperturbed equilibrium state) as a function of time for the fiducial spherical and Cartesian NFW runs with initial min($t_{\rm TI}/t_{\rm ff})=7.3$. 
 The red solid line shows the 
 expected linear growth for modes growing at the local thermal instability timescale (measured at $H$). The measured growth rates are 
 similar for spherical and Cartesian simulations, and comparable to the local growth rate. To have a significant linear regime, the maximum amplitude 
 of density perturbations in these runs is 0.01 (for other runs it is 0.3).}\label{fig:fig10} 
\end{figure}

Figure \ref{fig:fig10} shows the growth of density fluctuations at min($t_{\rm TI}/t_{\rm ff}$) for the fiducial NFW run (with smaller density perturbations) 
in spherical and Cartesian geometries. 
The perturbations in $0.9-1.1 H$ ($H$ corresponds to min$t_{\rm TI}/t_{\rm ff}$ in the equilibrium profile) grow roughly at the {\it local} thermal instability 
timescale ($t_{\rm TI}$; see Eq. \ref{eq:tTI}). More importantly, the exponential growth rate at early times is similar for spherical and plane-parallel atmospheres.
This is consistent with global linear stability analysis of section \ref{sec:global_linear}. Therefore, the apparent ease of condensation in a spherical atmosphere 
as compared to a plane-parallel one cannot be due to a difference in the linear growth rates. The non-linear saturated state shows larger fluctuations in a 
plane-parallel geometry because of large density perturbations propagating away from the midplane (see section \ref{sec:trig}). The amount of cold gas 
condensing (both in situ and anywhere) by 15 Gyr and the maximum radius/height till which cold gas exists ($r_{\rm cold, max}$) are similar for these 
small amplitude runs are similar to the fiducial run with $\approx 30$ times larger density perturbations (lister in Table \ref{table:sims}), implying that the 
magnitude of initial density perturbations (as long as $ \delta \rho/\rho < 1$)  does not affect the results over long times.

\subsubsection{Cold gas fraction: in situ versus condensation anywhere}

\begin{figure}
 \includegraphics[width=.5\textwidth]{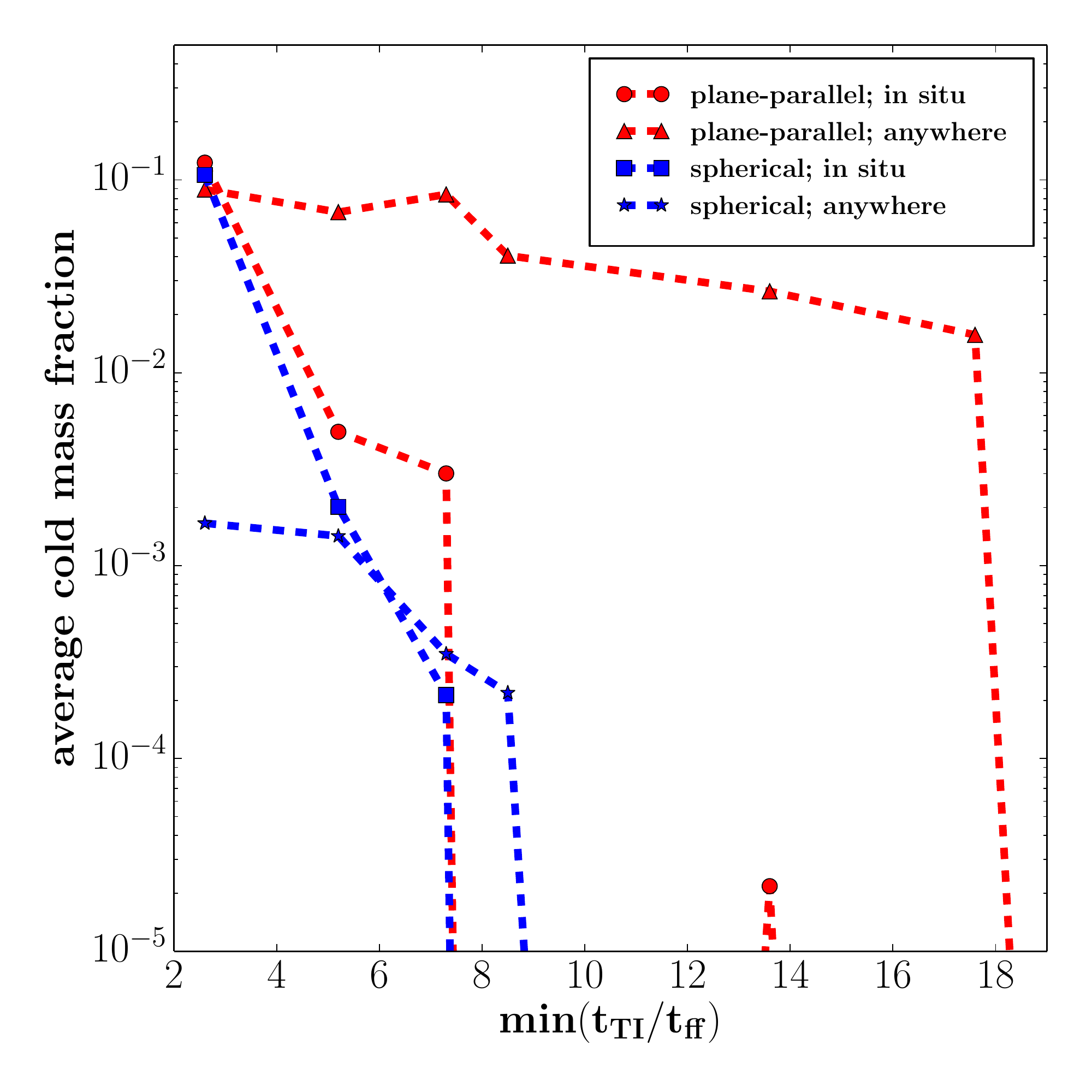}
 \caption{The ratio of average (over 15 Gyr) cold ($T<0.01$ keV) gas mass to the average total mass (in situ and anywhere) for various values of 
 initial min($t_{\rm TI}/t_{\rm ff}$) for the spherical and plane-parallel NFW runs. 
 While there is essentially no  cold gas  condensation for spherical runs with min$(t_{\rm TI}/t_{\rm ff}) > 8$, cold gas condenses out at smaller heights  
 for plane-parallel runs with min($t_{\rm TI}/t_{\rm ff}) \lesssim 18$. 
 }\label{fig:fig11}
\end{figure}

It is useful to distinguish cold gas which condenses {\it in situ} at min($t_{\rm TI}/t_{\rm ff}$) and cold gas which condenses {\it anywhere} in the box 
(typically within min[$t_{\rm TI}/t_{\rm ff}$]). It is sensible to make such a distinction both in theory (\citealt{mccourt12}) and 
observations (e.g., \citealt{mcdonald10}).
While centrally concentrated cold gas is associated with the interstellar medium of the central galaxy, extended cold gas is condensing  
out of the ICM now and falling toward the center (unless uplifted by jets or bubbles).

Figure \ref{fig:fig11} shows the average cold mass fraction, defined as the ratio of cold ($T<0.01$ keV) gas mass and the total gas mass between 1 kpc and 
$r_{\rm cold,max}$ (the largest radius, over all times, at which cold gas is seen), for several plane-parallel and spherical NFW simulations. 
The value of min$(t_{\rm TI}/t_{\rm ff})$ is tuned by changing $K_0$ in Eq. \ref{eq:ent}. The 
cold mass fraction is calculated as follows. First, we calculate $r_{\rm cold, max}$, the maximum radius out to which cold gas is detected in all the snapshots 
(more than 100 over the duration of the run). After that we calculate the average cold gas mass for all snapshots within $ 1~{\rm kpc} \leq  r \leq r_{\rm cold, max}$ 
($ 1~{\rm kpc}  \leq |z| \leq r_{\rm cold, max}$ for plane-parallel runs) and we divide it by the average total gas mass within the same volume. This way, we 
somewhat normalize for the fact that there is a much larger volume of dense gas for the plane-parallel setup.

Figure \ref{fig:fig11} shows that there is only a minor difference in the onset of cold gas condensation for spherical and plane-parallel simulations. For both, there is
negligible in situ (within $0.9-1.1 H$, where $H$ corresponds to the radius of min[$t_{\rm TI}/t_{\rm ff}$]) condensation for min$(t_{\rm TI}/t_{\rm ff})>8$. 
Cold gas does 
condense out at smaller radii for somewhat larger initial  min$(t_{\rm TI}/t_{\rm ff}) \approx 18$ in Cartesian geometry. However, there is no cold gas 
present anywhere for even larger min$(t_{\rm TI}/t_{\rm ff})$, corresponding to a lower density (see Table \ref{table:sims}). 
As expected, a more massive and extended cold gas condensation happens for the runs with a smaller min$(t_{\rm TI}/t_{\rm ff})$.
In both geometries, cold mass fraction over the whole box (as opposed to in situ condensation) falls slowly as a function of min$(t_{\rm TI}/t_{\rm ff})$. For 
both geometries, the critical $t_{\rm TI}/t_{\rm ff}$ for condensation anywhere in the box is within a factor of two of the critical value for in situ condensation.

\subsection{Triggered condensation}
\label{sec:trig}
\begin{figure}
 \includegraphics[width=.5\textwidth]{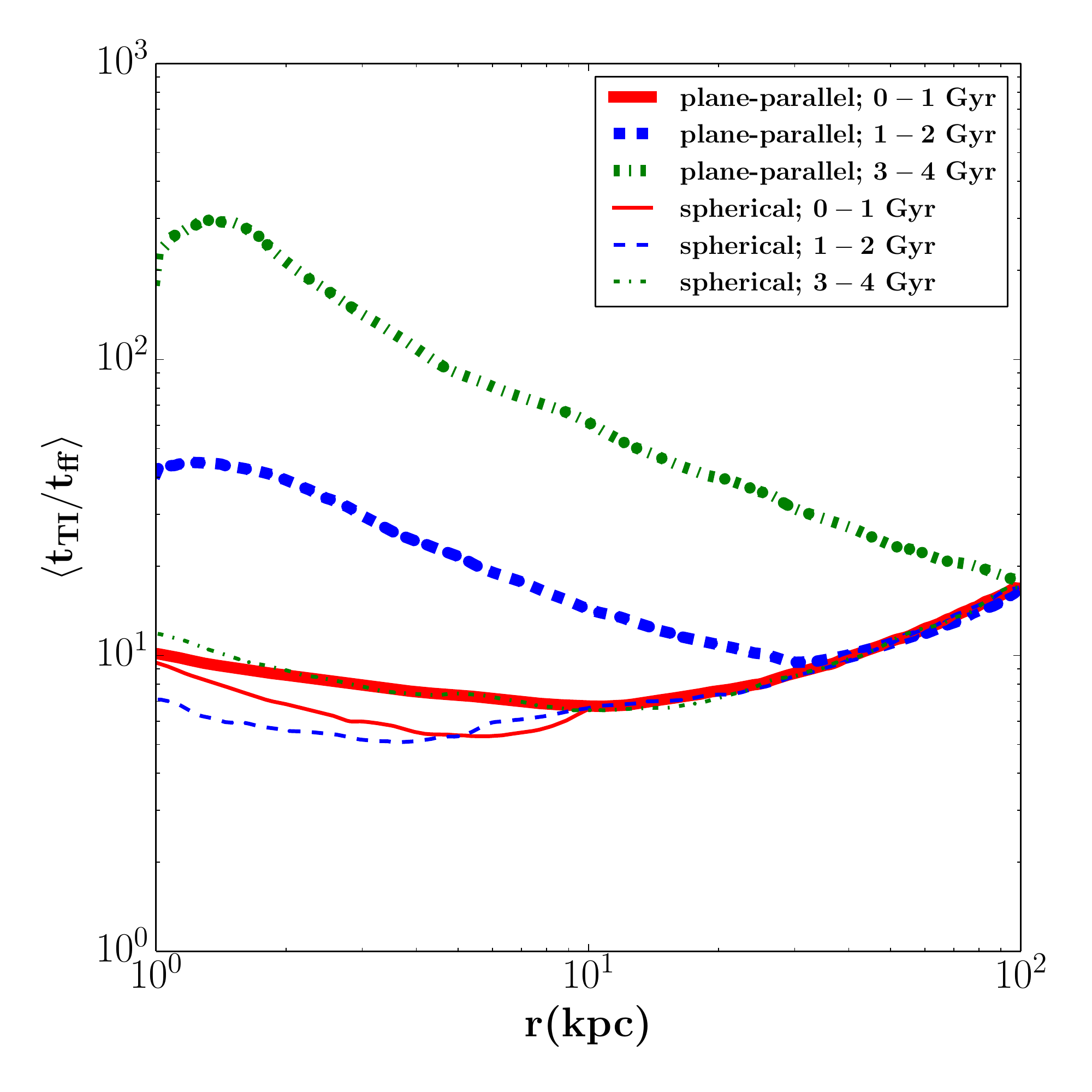}
 \caption{Evolution of shell-averaged $t_{\rm TI}/t_{\rm ff}$ (averaged over different times) for hot gas ($T>0.1$ keV) in fiducial spherical 
 and plane-parallel NFW runs. 
This ratio is much larger at late times for plane-parallel runs, implying that hot plumes move up and displace denser 
gas from there (see also Fig. \ref{fig:fig09}). In spherical atmosphere condensation
does not affect the overlying layers as much.} \label{fig:fig22}
\end{figure}

Figure \ref{fig:fig11} shows that condensation anywhere in the simulation box can occur for a larger value of min($t_{\rm TI}/t_{\rm ff}$) as 
compared to in situ condensation. Overdense blobs generated at min($t_{\rm TI}/t_{\rm ff}$) can fall in and induce large perturbations toward the 
center, and can lead to condensation there even if $t_{\rm TI}/t_{\rm ff}$ is larger than its value at the minimum.

Figure \ref{fig:fig09} shows a big difference in the state of the hot gas; namely, the hot gas density in plane-parallel simulation is much smaller than in the 
spherical run. To quantitatively study the state of the hot gas, in Figure \ref{fig:fig22} we plot the shell-averaged $t_{\rm TI}/t_{\rm ff}$ of the hot gas 
($T>0.1$ keV) at different times for the spherical and plane-parallel fiducial runs with initial min($t_{\rm TI}/t_{\rm ff}$)=7.3. For spherical simulations,
excess gas is removed from the hot phase in form of cold filaments, and the ICM settles with min($t_{\rm TI}/t_{\rm ff}) \gtrsim 10$, as also found in 
\citet{sharma12}. In contrast, the outer regions in the 
plane-parallel simulation get evacuated at later times due to plenty of hot bubbles rising from the innermost layers and displacing the denser gas. 
These rising low density gas plumes create large density perturbations ($\delta \rho/\rho \gtrsim 1$; see late times in Fig. \ref{fig:fig10}) 
and trigger further condensation (note that the min[$t_{\rm TI}/t_{\rm ff}$] threshold applies only for small density perturbations; cold gas can condense 
out from lower density atmospheres if density perturbations are large), explaining the 
presence of central cold gas for higher min($t_{\rm TI}/t_{\rm ff}$) in Figure \ref{fig:fig11} for plane-parallel simulations compared to spherical. The hot gas is
evacuated from most of the volume, with min($t_{\rm TI}/t_{\rm ff}$) at late times significantly larger than the critical value of 10, and cold gas settles 
down in the midplane (see Fig. \ref{fig:fig09}). Evacuation of gas away from the midplane is partly because of our imposed thermal balance in layers.
As gas cools through $10^7$ to $10^4$ K the rise in the cooling function results in excessive heating and thermal transport to upper layers.
In the realistic spherical atmosphere the outer layers are least affected by the 
tiny hot bubbles rising from the core region because the mass of cold gas condensing is much smaller.

\subsection{Condensation with different densities and gravities: plane-parallel runs}
\label{sec:diff_g}

While we have convincingly shown that geometry does not play a key role in determining condensation properties in our setup, we also want to assess the role
of gravitational acceleration and equilibrium entropy/density profiles in cold gas condensation. Various astrophysical coronae (e.g., accretion 
disk and galactic coronae; \citealt{das13,2012ApJ...745..148J,sharma12b}) have different distributions of gravitational accelerations and hot gas densities, 
and the details of cold gas formation may differ. 

\begin{figure}
 \includegraphics[width=.5\textwidth]{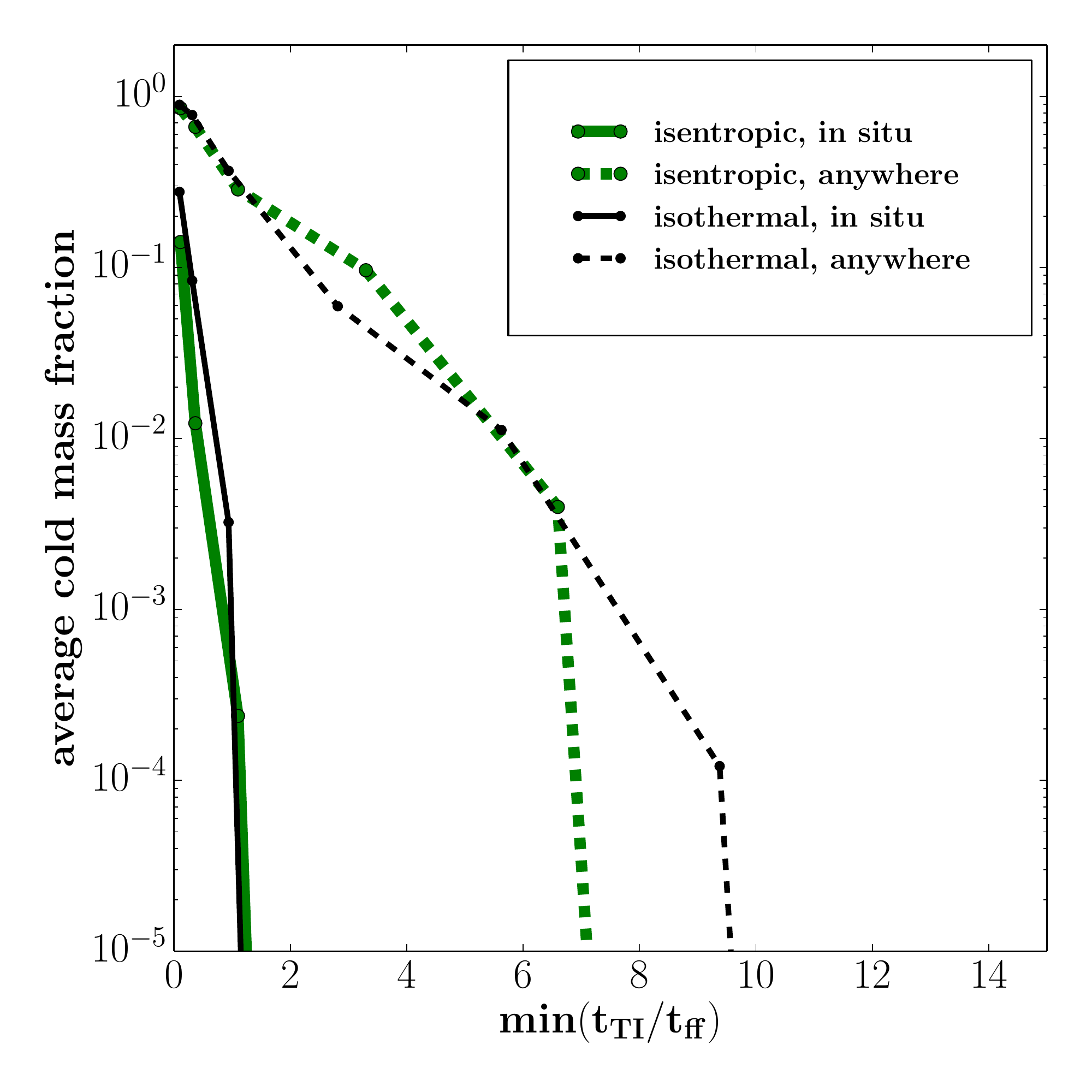}
 \caption{The ratio of average (over 180 code units) cold mass ($T\leq T_0/10$) to the average total mass as a function of min($t_{\rm TI}/t_{\rm ff}$) for 
 idealized isothermal/isentropic Cartesian MSQP runs. The average cold mass fraction, for both in situ ($0.9-1.1 H$) and anywhere ($|z| <0.1$ is excluded)
  condensation, is qualitatively similar to Fig. \ref{fig:fig11} for NFW runs. However, the $t_{\rm TI}/t_{\rm ff}$ threshold for condensation anywhere 
  is a factor of $\approx 7$ larger than in situ condensation. } \label{fig:fig12}
\end{figure}

In order to compare with \citet{mccourt12}, Figure \ref{fig:fig12} shows the in situ and anywhere cold mass fraction for idealized isothermal and 
isentropic MSQP runs in Cartesian 
geometry. Both gravity
and density/temperature profiles are different for these runs as compared to the NFW runs (see Fig. \ref{fig:fig07}). 
However, the results are qualitatively similar; there is a cut-off 
min($t_{\rm TI}/t_{\rm ff}$) beyond which no cold gas condenses, either in situ at min($t_{\rm TI}/t_{\rm ff}$) (critical value $\approx 1$) or anywhere 
in the simulation domain (critical value $\approx 8$). A smaller threshold of $t_{\rm TI}/t_{\rm ff}$ for idealized MSQP runs (especially for in situ condensation) 
as compared to NFW runs can be understood from the shapes of $t_{\rm TI}/t_{\rm ff}$ profiles in Figure \ref{fig:fig07}. Going toward the center, 
this ratio rises more steeply for idealized MSQP runs as 
compared to NFW runs. This means that there is a much larger volume prone to condensation for NFW atmospheres in contrast to the idealized MSQP 
ones, leading to the relative ease of condensation in former. There is no significant difference between condensation in isothermal and isentropic 
idealized atmospheres, implying that background entropy stratification does not play a critical role in nonlinear evolution.

 \begin{figure}
 \includegraphics[width=.5\textwidth]{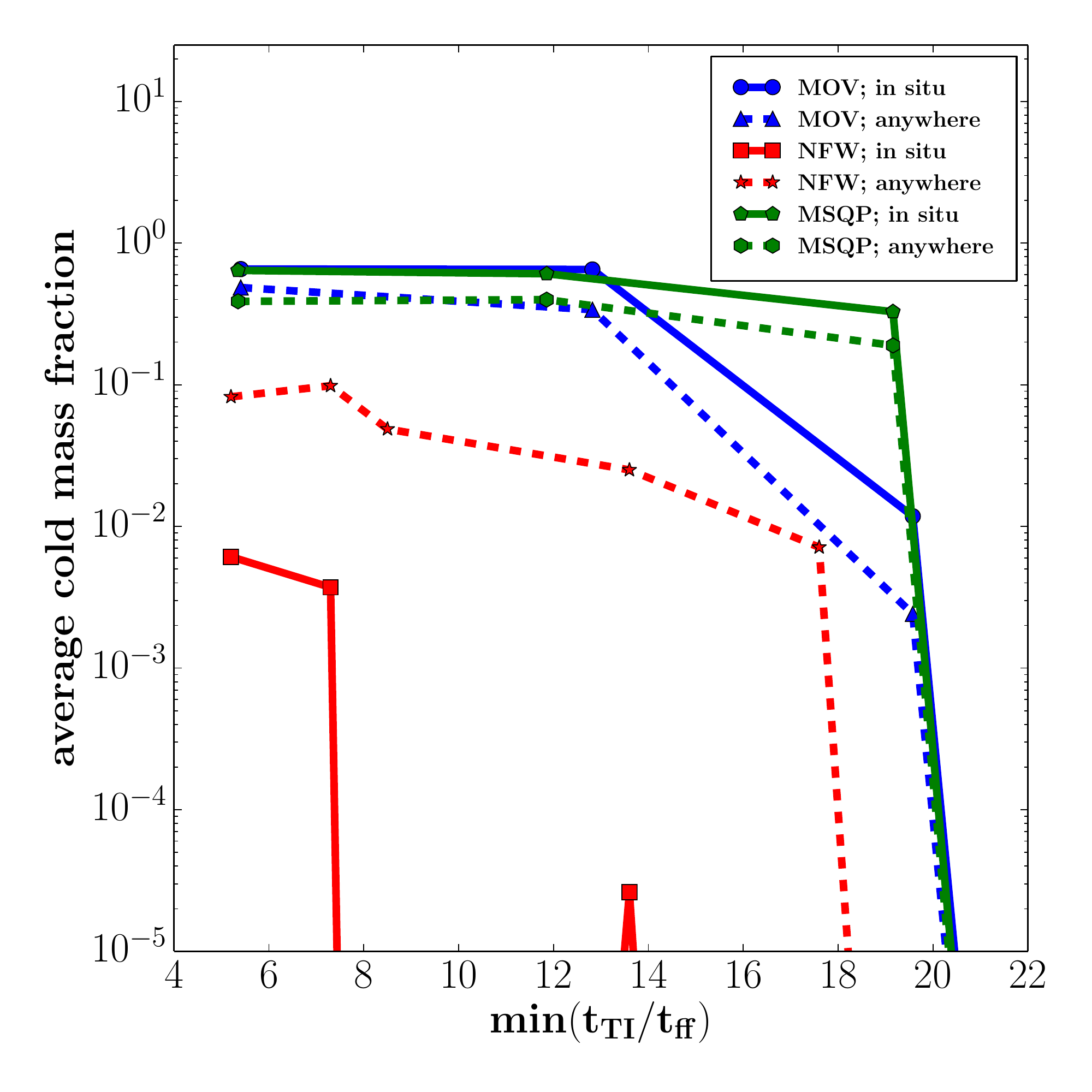}
 \caption{Average cold gas mass fraction (over 10.9 Gyr) for plane-parallel MOV and NFW runs (also shown are results from MSQP runs averaged over 
 6.5 Gyr) for different initial min($t_{\rm TI}/t_{\rm ff}$). While the in situ and anywhere condensation ratio is very similar for MOV and MSQP runs, the 
 threshold min($t_{\rm TI}/t_{\rm ff}$) for in situ condensation in NFW runs is a factor of about two smaller as compared to condensation anywhere in the 
 box. Note that unlike NFW runs, in situ cold mass fraction is larger than anywhere cold mass fraction for MOV/MSQP runs.}
 \label{fig:fig13}
\end{figure}

Figure \ref{fig:fig13} shows the average cold mass fraction (both in situ and anywhere) as a function of min($t_{\rm TI}/t_{\rm ff}$) for similar initial entropy 
profiles (Eq. \ref{eq:ent}) but for NFW and MOV/MSQP gravities (Eqs. \ref{eq:g_NFW}, \ref{eq:g_MOV}/\ref{eq:g_MSQP}). The noticeable differences 
between the NFW and MOV/MSQP gravities
are that the amount of cold gas condensing is higher for MOV/MSQP, and that the threshold for in situ and anywhere condensation are similar for MOV/MSQP 
but differ by about a factor of two for NFW.  In fact, the in situ cold fraction is higher for MOV/MSQP profiles that have min($t_{\rm TI}/t_{\rm ff}$) at the center. 
For profiles with min($t_{\rm TI}/t_{\rm ff}$)  away from the center (e.g., NFW), cold gas condensation is
seeded far away from the center but eventually happens much closer in. For min($t_{\rm TI}/t_{\rm ff}$) at the center, cold gas condensation is anyway 
most efficient at the center.

\begin{figure}
 \includegraphics[width=.5\textwidth]{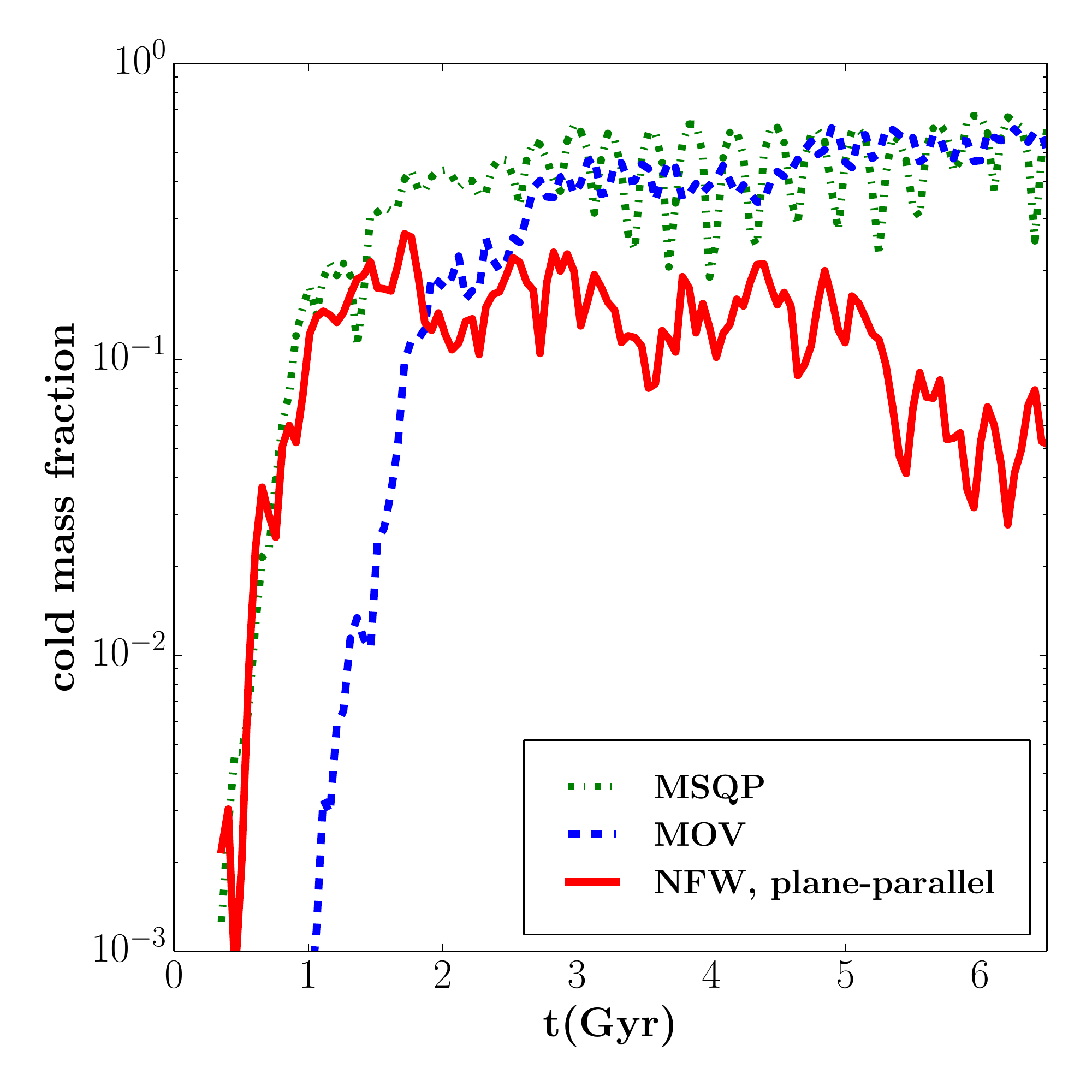}
 \caption{Comparison of onset time and cold mass fraction as a function of time for Cartesian runs with NFW (with equilibrium 
 min$[t_{\rm TI}/t_{\rm ff}]=5.2$), MOV (min$[t_{\rm TI}/t_{\rm ff}]=5.4$) and MSQP (min$[t_{\rm TI}/t_{\rm ff}]=5.3$) gravities. Cold mass fraction is
 higher for MOV and MSQP runs as compared to the NFW run, which also shows a slow decline with time.}\label{fig:fig14}
\end{figure}

Figure \ref{fig:fig14} shows cold gas mass fraction as a function of time for Cartesian runs with NFW, MOV, and MSQP gravities with initial
min$(t_{\rm TI}/t_{\rm ff})\approx 5$; the corresponding $t_{\rm TI}/t_{\rm ff}$ and density profiles are show in Figure \ref{fig:fig07}. Figure \ref{fig:fig07}
shows that the central density of MSQP model is the largest. Therefore, cold gas condenses out earliest in this model. The $t_{\rm TI}/t_{\rm ff}$ profiles 
are quite similar
for MSQP and MOV models. Cold gas condenses out almost 
simultaneously for MSQP and NFW models because of a similar density (and hence the cooling time) at the location of min($t_{\rm TI}/t_{\rm ff}$). While the cold
gas fraction (also the cold gas mass) is higher and attains an almost constant level for MOV and MSQP models in Figure \ref{fig:fig14}, it is smaller and decreases gradually for the NFW model. This is very likely
due to the qualitatively different $t_{\rm TI}/t_{\rm ff}$ profiles for NFW as compared to MOV and MSQP atmospheres. With a $t_{\rm TI}/t_{\rm ff}$
minimum away from the center, the gas close to the center is not prone to multiphase cooling.

The comparison of average cold fraction, both in situ and anywhere, for MSQP gravity in Figures \ref{fig:fig12} 
and \ref{fig:fig13} shows that the condensation properties are essentially a function of the value and shape of $t_{\rm TI}/t_{\rm ff}$ profiles (and not just gravity 
or density profiles). Profiles with a
minimum far away from the center (NFW and idealized MSQP; see Fig. \ref{fig:fig07}) show lower amount of cold gas and a larger difference between the 
critical threshold for in situ and anywhere condensation. 

\subsection{Enhanced condensation in midplane with no gravity}
\label{sec:midplane_cooling}

\begin{figure*}
 \includegraphics[width=.95\textwidth]{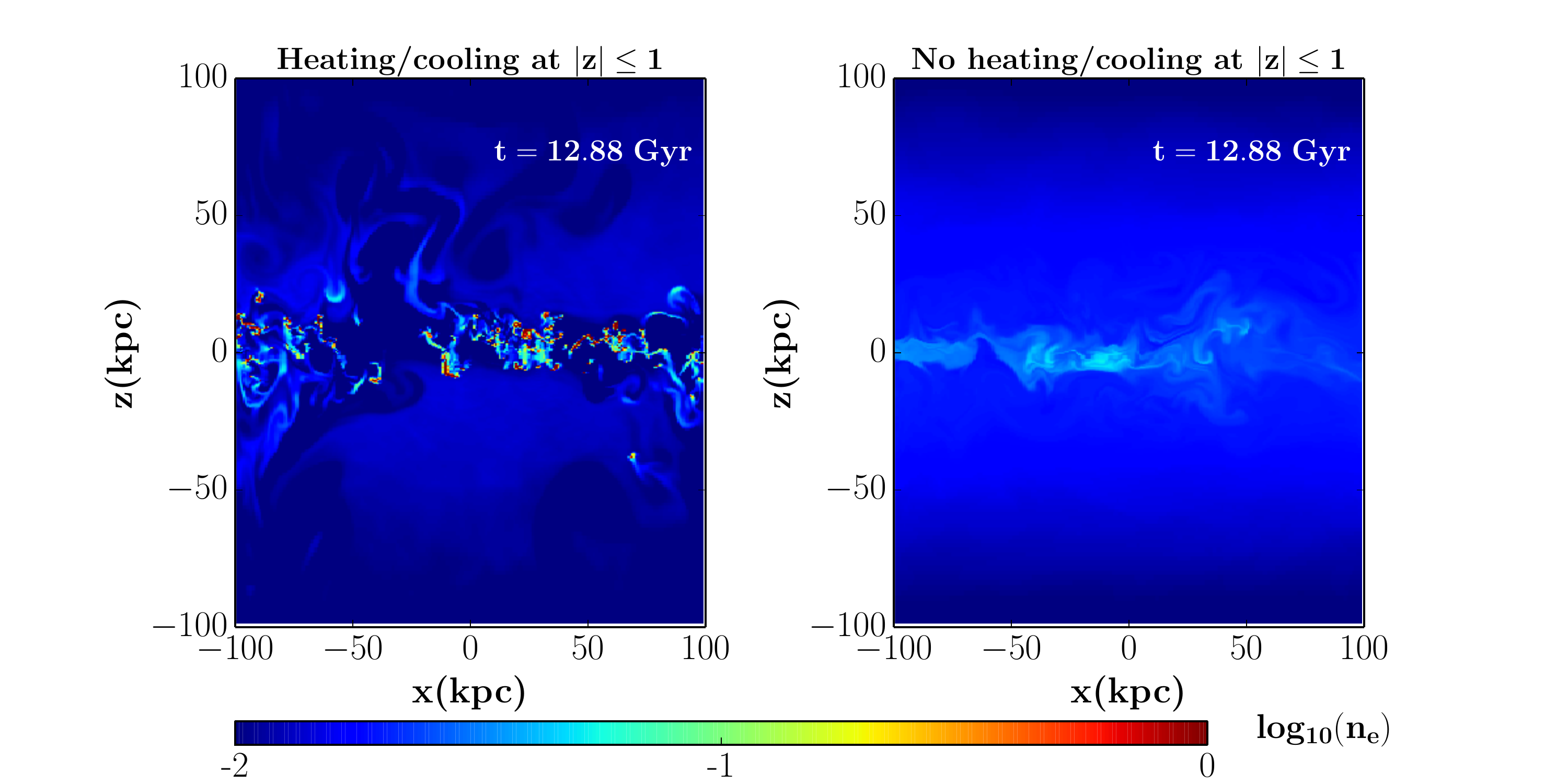}
 \caption{Electron number density snapshots for MOV Cartesian runs with (left panel) and without (right panel) cooling/heating in the midplane ($|z|<1$).
 There is widespread multiphase gas with heating and cooling allowed in the midplane.}\label{fig:fig15}
\end{figure*}

Based on their simulations, \citet{meece15} have recently argued that  there is no threshold of $t_{\rm TI}/t_{\rm ff}$ for condensation to occur. 
That is, condensation will always occur, provided one waits long enough. This conclusion differs from the claim of \citet{mccourt12}, 
who turn off cooling and heating at the midplane of the simulation domain. While $t_{\rm TI}/t_{\rm ff}$ for MOV and MSQP gravities 
formally has a finite value as $z \rightarrow 0$, gas in thermal balance at the midplane is essentially unaffected 
by gravity and will become multiphase over a cooling time for all cases. This is what leads to the formation of substantial quantities of cold gas in the 
simulations of  \citet{meece15} even if min($t_{\rm TI}/t_{\rm ff} \gtrsim 30$).

Figure \ref{fig:fig15} shows a comparison of the Cartesian MOV simulations, with an initial min($t_{\rm TI}/t_{\rm ff}$)=26.2, with (right panel) and without (left panel)
 heating/cooling turned off in the midplane ($|z| < 1$ kpc). A comparison of the two panel clearly shows that there is much more cold gas condensing out in
 the latter case. Condensation occurs even beyond the midplane; this is seeded by perturbations generated at the midplane and by cold gas rising up 
 due to buoyant hot bubbles. Table \ref{table:sims} also shows a much larger cold gas fraction for the run that allows cooling and heating in the 
 midplane. Thus, turning off multiphase gas production in the midplane is necessary to carefully assess the source of extended cold gas in cool cluster cores.

\section{Conclusions}
\label{sec:conc}

We have used global linear analysis and idealized numerical simulations in global thermal balance to address the question of cold gas condensation in
cool cluster cores and other astrophysical coronae. We have not varied all the parameters in our simulations because earlier works had explored those 
(e.g., a comparison of 2-D and 3-D and the effect of small initial density perturbations). Particular attention is paid to the influence of geometry and the 
profile of the ratio of the cooling 
time to the free-fall time ($t_{\rm cool}/t_{\rm ff}$). We have used different forms of gravities (Fig. \ref{fig:fig01}) and hydrostatic atmospheres 
(Figs. \ref{fig:fig05} \& \ref{fig:fig07}) to carry out our investigations. Previous works (except for the recent paper of \citealt{meece15}) somewhat 
erroneously assumed 
that it was easier for cold gas to condense out in spherical geometry as compared to a plane-parallel atmosphere. The key reason for this confusion
was that \citet{mccourt12} focused on in situ condensation of cold gas but \citet{sharma12} quantified condensation 
of cold gas anywhere in the box, not necessarily at the location of the minimum in $t_{\rm cool}/t_{\rm ff}$.
 
 The main conclusions of our paper are:
 
 \begin{itemize}
 
 \item {\it Linear growth rate of global modes is independent of geometry and gravity:} Global linear analysis shows that the growth rate of the global 
 modes is the same for spherical and Cartesian geometries (Fig. \ref{fig:fig06}). This is also verified in numerical simulations by measuring the density growth 
 in the two geometries (Fig. \ref{fig:fig10}). Moreover, the growth rate of global modes is similar to the local 
 isobaric thermal instability growth rate. Therefore, any differences observed for condensation in different geometries and gravities are due to nonlinear 
 effects such as seeding of condensation due to rising/sinking underdense/overdense blobs.
 
  \item {\it Critical min($t_{\rm TI}/t_{\rm ff}$) is essentially independent of geometry but much higher cold gas condenses in plane-parallel geometry:} 
 Cold gas condensation, in our simulations with small density perturbations, occurs only if $t_{\rm TI}/t_{\rm ff}$ is smaller than a critical value close to 10.
  Our
  plane-parallel NFW simulations show that much more cold gas condenses out of the hot phase compared to spherical simulations (Fig. \ref{fig:fig09}). 
  This is because of the much bigger volume which is prone to multiphase 
  condensation for the former. A large amount of multiphase gas leads to strong heating (because of thermal balance ansatz), and large fluctuations lead to the condensation
  of a lot of gas, even in regions where $t_{\rm TI}/t_{\rm ff} > 10$. For realistic spherical runs, on the other hand, only a small volume is prone to condensation 
  and the fluctuations induced due to this are not sufficient to affect the overlying gas with $t_{\rm TI}/t_{\rm ff} > 10$ (right panels of Fig. \ref{fig:fig09}).
  The {\it onset} of cold gas condensation, both in situ where $t_{\rm TI}/t_{\rm ff}$ is minimum and anywhere in the box,  happens at a similar value of 
  threshold min($t_{\rm TI}/t_{\rm ff}$) for spherical and plane-parallel setups (Fig. \ref{fig:fig11}).
  
 \item {\it Role of $t_{\rm TI}/t_{\rm ff}$ profile:} The gravity and gas density profiles, via the dimensionless ratio $t_{\rm TI}/t_{\rm ff}$ (the ratio of local 
 thermal instability timescale and the free-fall time), play a crucial role in governing cold gas condensation. In cases where min$(t_{\rm TI}/t_{\rm ff})$ is far away 
 from the center (e.g., NFW, idealized MSQP; see Fig. \ref{fig:fig07}), the min$(t_{\rm TI}/t_{\rm ff})$ threshold for in situ condensation is 
 considerably smaller compared to the threshold for condensation 
 anywhere (Figs. \ref{fig:fig11}, \ref{fig:fig12}). For MOV and MSQP profiles (Fig. \ref{fig:fig07}), for which  min$(t_{\rm TI}/t_{\rm ff})$ 
 lies at the center, there is essentially no difference between in situ and condensation anywhere (see Fig. \ref{fig:fig13}). For idealized MSQP 
 and NFW profiles, with 
 min$(t_{\rm TI}/t_{\rm ff})$ away from the center, overdense gas falling in from min($t_{\rm TI}/t_{\rm ff}$) triggers (nonlinearly) condensation toward the center 
 where $(t_{\rm TI}/t_{\rm ff})$ is larger. Atmospheres with weaker gravity at the center, such as MOV (see Fig. \ref{fig:fig01}), 
 allow fluctuations arising at the center (where 
 there is multiphase cooling) to larger heights where they can trigger cooling, leading to multiphase condensation at greater heights ($r_{\rm cool, max}$ is 
 larger for MOV runs in Table \ref{table:sims}).
 
 \item {\it Triggered multiphase gas due to large density perturbations:} Large density perturbations can give rise to the condensation of multiphase gas 
 even if $t_{\rm cool}/t_{\rm ff}$ is larger than the critical value (e.g., Fig. 7 in \citealt{ashmeet14}). This is what happens at the center of NFW runs which
 do not show in situ condensation at min($t_{\rm cool}/t_{\rm ff}$)  but toward the center. Large overdensites generated at min($t_{\rm cool}/t_{\rm ff}$) 
 propagate inward and lead to cold gas condensation at smaller radii. Similarly, for plane-parallel runs, density fluctuations originating close to 
 min($t_{\rm cool}/t_{\rm ff}$) propagate outwards and lead to widespread dropout as compared to a realistic spherical geometry of the ICM. 
 A dramatic example of triggered 
 multiphase condensation is shown in Figure \ref{fig:fig15} which compares two MOV plane-parallel simulations in one of which cooling/heating is allowed 
 in the midplane. Since the gas in the midplane does not experience any gravity, cold gas condenses out in the midplane and the large density perturbations
 generated there trigger condensation far away from the center. \citet{meece15} do not turn off heating and cooling at the center and therefore their results
 are strongly affected by condensation triggered due to density fluctuations propagating away from the midplane. This is the reason they do not see a 
 min($t_{\rm cool}/t_{\rm ff}$) threshold for condensation in their simulations. 

 \end{itemize}

Our key result is
that cold gas condensation, starting with small amplitude perturbations, is expected only if min($t_{\rm TI}/t_{\rm ff}) \lesssim 10$. This results is only weakly
sensitive to the gravitational potential and density/temperature profiles. It is not possible to understand this threshold from linear physics arguments. 
Nonlinear effects, such as triggering of multiphase condensation due to motion of overdense/underdense blobs across layers, play important role 
in explaining the results from various idealized numerical simulations.

\section*{Acknowledgements}
This work is partly supported by the DST-India grant no. Sr/S2/HEP-048/2012 and an India-Israel joint research grant (6-10/2014[IC]). 
We thank Matt Kunz for sharing the code that was used in  \citet{Latkunz2012}. We also thank Ramesh Narayan for useful discussions 
on the global linear stability analysis.

\bibliographystyle{mn2e}
\bibliography{bibtex}

\newpage
\appendix
\section{Global linear equations in spherical coordinates}
\label{app:linear}
The perturbed quantities in Eqs. \ref{eq:eq4}-\ref{eq:eq8} can be expanded using a spherical harmonic basis as follows,
$$
\rho_1 = R_\rho Y_l^m,
p_1 = R_p Y_l^m,
s_1 = R_s Y_l^m,
T_1=R_T Y_l^m,
$$
$$
v_{r1} = R_r Y_l^m,
v_{\theta 1} = R_\theta  \frac{\partial Y_l^m(\theta,\phi)}{\partial \theta},
v_{\phi 1} =  \frac{R_\phi}{\sin \theta} \frac{\partial Y_l^m(\theta,\phi)}{\partial \phi},
$$
where $Y_l^m(\theta,\phi)$ is the spherical harmonic of order $(l,m)$ and $R$ carries the radial dependence. These forms 
are obtained by comparing the $r,~\theta,~\phi$ dependence of various terms in Eqs. \ref{eq:eq4}-\ref{eq:eq8}, which also imply 
that $R_\phi=R_\theta$. The equations become
\ba
\label{eq:ee1}
\sigma R_\rho &=& - \frac{1}{r^2} \frac{d}{dr} (r^2 \rho_0 R_r) + l(l+1) \frac{\rho_0 R_\theta}{r},\\
\label{eq:ee2}
\sigma R_r &=&   -\frac{1}{\rho_0} \frac{d}{dr} \left [ p_0 \left ( \frac{R_s}{s_0} + \gamma \frac{R_\rho}{\rho_0} \right ) \right ]-\frac{gR_\rho}{\rho_0}, \\
\label{eq:ee3}
\sigma R_\theta &=& -\frac{p_0}{r \rho_0} \left ( \frac{R_s}{s_0} + \gamma \frac{R_\rho}{\rho_0} \right ), \\
\label{eq:ee4}
\sigma R_s &=& \frac{\gamma s_0 N^2}{g} R_r \\
\nonumber
&-& \frac{s_0}{t_{\rm cool 0}} \left [ 2 \frac{R_\rho}{\rho_0} + \frac{d\ln \Lambda}{d\ln T} \left( \frac{R_s}{s_0} + (\gamma-1) \frac{R_\rho}{\rho_0} \right) \right].
\nonumber
\ea 

The local forms (assuming $e^{i k_x x}$ dependence; $x$ is the local coordinate along $\theta$) of Eqs.\ref{eq:eq4} - \ref{eq:eq8} are
\ba
\label{eq:eq9}
&& \sigma \rho_1 + \frac{1}{r^2} \frac{\partial}{\partial r}  ( r^2 \rho_0 v_{r1} ) + ik_x \rho_0 v_{x 1} = 0, \\
\label{eq:eq10}
&& \sigma \rho_0 v_{r1} = - \frac{\partial p_1}{\partial r} - \rho_1 g, \\
\label{eq:eq11}
&& \sigma \rho_0 v_{x 1} = -ik_x p_1, \\
\label{eq:eq13}
&& \sigma \frac{s_1}{s_0}  +  \frac{\gamma N^2 v_{r1}}{g} = \frac{-1}{t_{\rm cool 0}} \left [ 2 \frac{\rho_1}{\rho_0} 
+ \frac{d \ln \Lambda}{d \ln T} \frac{T_1}{T_0}  \right ].
\ea
Both the local (in $\theta$) and global equations can be solved as eigenvalue problems with given parameters $l,~m$ (for global setup) or 
$k_x$ (for local setup). For plane-parallel (as opposed to spherical) global linear analysis we just have to set the second term in Eq. \ref{eq:eq9}
to be $d (\rho_0 v_{z1})/dz$.

\section{Pseudospectral method}
\label{app:pseudo}
We use the pseudospectral method (chapters 6 \& 7 in \citealt{boyd2000}; \citealt{Latkunz2012}) to solve the set of eigen-equations, Eqs. \ref{eq:eq9}-\ref{eq:eq13} 
(Eqs. \ref{eq:ee1}-\ref{eq:ee4} for spherical global $\theta-\phi$ modes), by approximating the eigenfunctions in terms of $n$ Chebyshev 
polynomials of first kind ($T$) as:
\ba
\nonumber
&& \frac{\rho_1}{\rho_0} = \sum_{j=0}^{n-1} {\tilde{\rho}_j}{T_j{\left( \xi \right)}}; 
\frac{v_{r1}}{c_0} = \sum_{j=0}^{n-1} {\tilde{v}_{rj}}{T_j{\left( \xi \right)}}; \\
&& \frac{v_{x 1}}{c_0} = \sum_{j=0}^{n-1} {\tilde{v}_{x j}}{T_j{\left(\xi \right)}};
\frac{s_1}{s_0} = \sum_{j=0}^{n-1} {\tilde{s}_j}{T_j{\left(\xi \right)}};
\label{eq:exp}
\ea
where $\xi = 2(r - r_{\rm in})/(r_{\rm out}-r_{\rm in}) - 1$ maps the domain going from $r_{\rm in}$ to $r_{\rm out}$ on to $-1 \leq \xi \leq 1$, as 
Chebyshev polynomials are complete and othonormal only over this domain.

The radial domain is partitioned into a Gauss-Lobatto grid so that $\xi(i)$ can be generated in the following way
$$
\xi[i] = \cos\left(\frac{i \pi}{n-1}\right); 
r[i] = (\xi[i] +1)\frac{(r_{\rm out} - r_{\rm in})}{2} + r_{\rm in},
$$
where $i= 0,..,(n-1)$ denotes the $n$ G-L points.
At each $r[i]$, each of the equilibrium profiles of $\rho_0$, $p_0$, $s_0$, $T_0$, $N^2$ and $t_{\rm cool 0}$ are obtained and we formulate the 
$4n \times 4n$ (Eqs. \ref{eq:ee1}-\ref{eq:ee4}) matrix eigenvalue problem; some of the rows of the matrix incorporate our boundary 
conditions (see section \ref{sec:global_linear}). Incorporating the above into the differential eigen-system (Eqs. \ref{eq:ee1}-\ref{eq:ee4} for global spherical 
modes and Eqs. \ref{eq:eq9}-\ref{eq:eq13} for local modes), we can convert it into a matrix form so that we get a matrix whose eigenvalues ($\sigma$)
 are the growth rates (or decay rates if real parts are negative; the imaginary parts correspond to oscillation frequencies).

The discretized eigenproblem takes the form (assuming that the repeated index is summed over all possible values)
 \be
 \label{eq:mat}
 {\cal L}_{ij} v_j = \sigma {\cal M}_{ij} v_j
 \ee
 where ${\cal L}$ and ${\cal M}$ are $4n \times 4n$ matrices; $i$ is the radial grid index for dependent variables ($\rho_1,~v_{r1}$, etc.) 
 and $j$ stands for the order of Chebyshev basis for $\rho_1,~v_{r1}$, etc. 
 As a result, we can consider
 ${\cal L}$ and ${\cal M}$ to be consisting of $4\times4$ blocks, each of size $n \times n$. For example, in terms of our dependent variables, Eq. \ref{eq:eq11} 
 can be written as $\sigma \rho_0 v_{x 1} =  -i k_x p_0 (\gamma \rho_1/\rho_0 + s_1/s_0)$. The discretized equations for $v_{x 1}$ 
 correspond to rows from $2n$ to $3n-1$ of Eq. \ref{eq:mat}; rows $2n$ and $3n-1$ implement the outer and inner boundary conditions. 
 For Eq. \ref{eq:eq11}, Matrix ${\cal L}$ will have non-zero entries in columns $j=0,..,n-1$ (corresponding to $\rho_1$) and $j=3n,..,4n-1$ (corresponding to 
 $s_1$), and ${\cal M}$ has non-zero entries in columns $j=2n,..,3n-1$. Once we obtain the eigenvalues and eigenvectors, we substitute them in Eq. \ref{eq:exp}
 to obtain the spatial profiles of eigenmodes. Since our equations and boundary conditions are homogeneous, the amplitude of modes is arbitrary within a 
 multiplicative factor and the real and imaginary components of eigenmodes and eigenvalues can be combined with an arbitrary phase and amplitude, e.g.,
 $$
 \rho_1 =  K \left ( \rho_{1r} \sin \phi + \rho_{1i} \cos \phi \right ),
 $$
 where $K$ and $\phi$ are arbitrary real numbers and $\rho_{1r/i}$ is the real/imaginary part of the density eigenmode. The other complex eigenmodes (e.g., $v_{r1r}$, $v_{r1i}$) are combined with the same value of $K$ and $\phi$.

As mentioned in \citealt{boyd2000}, the pseudospectral method gives several spurious eigenvalues and one must perform convergence studies to choose
the physically correct modes (e.g., Fig. \ref{fig:fig02}). Also, the higher wavenumber modes are better resolved as we increase the resolution ($n$).
 
\label{lastpage}

\end{document}